\immediate\write18{makeindex \jobname.nlo -s nomencl.ist -o \jobname.nls}
\documentclass[journal]{IEEEtran}
\usepackage{times}
\usepackage{graphics}
\usepackage{subfigure}
\usepackage{epsfig}
\usepackage{stfloats}
\usepackage{multirow}
\usepackage{multicol}
\usepackage{booktabs}
\usepackage{footnote}
\usepackage{arydshln}
\usepackage{mathptmx}
\usepackage{amsmath}
\usepackage{amssymb}
\usepackage{amsfonts}
\usepackage{yhmath}
\usepackage[noend]{algpseudocode}
\usepackage{algorithmicx,algorithm}
\usepackage{ifthen}
\usepackage[noprefix]{nomencl}
\makenomenclature
\usepackage{xcolor}
\usepackage[colorlinks,
        linkcolor=black,
        anchorcolor=black,
        citecolor=black,
        urlcolor=black
        ]{hyperref}
\usepackage{cite}

\begin{document}

    \title{Frequency Stability-Constrained Unit Commitment: \\Tight Approximation using Bernstein Polynomials}

    \author{
        Bo Zhou, \IEEEmembership{Member, IEEE,}
        Ruiwei Jiang, \IEEEmembership{Member, IEEE,} and
        Siqian Shen
        \vspace{-1em}

        \thanks{
            Ruiwei Jiang was supported in part by the U.S. National Science Foundation under Grant ECCS-1845980.
            Bo Zhou and Siqian Shen were supported in part by the U.S. Department of Energy, Grant \#DE-SC0018018.
            (Corresponding author: Bo Zhou, email: bozum@umich.edu)

            The authors are with the
            Department of Industrial and Operations Engineering, University of Michigan, Ann Arbor, MI 48109, USA.
        }
    }

    \maketitle

    \begin{abstract}
        As we replace conventional synchronous generators with renewable energy, the frequency security of power systems is at higher risk.
        This calls for a more careful consideration of unit commitment (UC) and primary frequency response (PFR) reserves.
        This paper studies frequency stability-constrained UC under significant wind power uncertainty.
        We coordinate the thermal units and wind farms to provide frequency support, wherein we optimize the variable inverter droop factors of the wind farms for cost minimization.
        In addition, we adopt distributionally robust chance constraints (DRCCs) to handle the wind power uncertainty.
        To depict the frequency dynamics, we incorporate a differential-algebraic equation (DAE) with dead band into the UC model.
        Notably, we apply Bernstein polynomials to derive tight inner approximation of the DAE and drive a mixed-integer linear representation, which can be solved by off-the-shelf solvers.
        Case studies demonstrate the tightness and effectiveness of the proposed method in guaranteeing frequency security.
        \vspace{-1ex}
    \end{abstract}


    \vspace{-1ex}
    \section{Introduction}
        \subsection{Background}
        \IEEEPARstart{T}{he} energy and climate crisis has motivated sustainable development and an explosive increase of renewable energy use worldwide \cite{Irena-2022}.
        However, the significant integration of renewables brings up many challenges.
        Among them, frequency security has drawn increasing attention~\cite{XieCarvalho-4359}.
        Different from synchronous thermal units that have rotors and governors, renewable generation is usually controlled by inverters, which cannot provide rotational inertia and primary frequency response (PFR) under the typical maximum power point tracking mode~\cite{TaftiWang-4365}.
        As renewables gradually replace thermal units, the total rotational inertia and PFR capacity in the power system decrease.
        Consequently, when a sudden power imbalance occurs, the system frequency will change more sharply and deviate from the nominal frequency more drastically, threatening the frequency security.

        A significant amount of effort has been devoted to ensuring frequency security under a high renewable penetration.
        Straightforwardly, one can keep sufficient synchronous units online, yet this may induce large curtailment of renewables.
        Some prior studies considered emergency resources for frequency support, such as energy storage~\cite{AbiariDaemi-5677} and HVDC links~\cite{ZhuBooth-4362}.
        Especially, modern energy storage, such as flywheel \cite{SaberiOskoueeKamali-5664} and battery \cite{WenLi-4681}, plays an important role in supporting frequency stability of renewable-integrated power systems.
        However, these approaches incur either high investment cost or operational risks.
        Fortunately, modern development of control technologies has allowed renewables themselves to contribute to frequency support~\cite{RocabertLuna-4363}.
        Using proper control strategies, renewables can adjust their integrated power according to frequency deviation and provide virtual inertia and PFR capacity~\cite{DehghanitaftiKonstantinou-4360}.
        This is relatively cheaper and more practical as the integration of renewables further increases.
        As a result, renewable generation is becoming an important resource that can be coordinated with thermal units to guarantee frequency security.

        Nevertheless, providing frequency support changes operations in power systems.
        On the one hand, sufficient inertia is required to depress sharp frequency variations, and hence, more units may need to be put online~\cite{RestrepoGaliana-4679, GolpîraAmini-5270}.
        On the other hand, sufficient PFR reserves are required for effective real-time PFR, for which thermal units need to reserve some generation capacity~\cite{GalianaBouffard-4678} and renewable generation will be operated under the deloaded mode~\cite{VidyanandanSenroy-4364}.
        Both of the required changes will increase the operational cost, yet they are coupled to influence frequency dynamics and further affect frequency security.
        Therefore, it is of great significance to carefully decide unit commitment (UC) and PFR reserves, in order to maintain frequency security in an economic manner.

        \vspace{-2ex}
        \subsection{Literature Review}
        A significant amount of literature has considered frequency security in scheduling, such as economic dispatch~\cite{LeeBaldick-4682}, UC~\cite{BadesaTeng-4680}, and energy market~\cite{LiGuo-4669}.
        However, due to the complex coupling of frequency dynamics, various frequency support resources, and the uncertainty of renewables, it remains extremely challenging to compute frequency stability-constrained UC accurately and tractably, which is the research focus of this paper.
        In the following, we conduct the literature review from two aspects, modeling frequency security constraints and handling uncertainties for frequency security.

        Three metrics of frequency dynamics are often considered, including the rate of change of frequency (RoCoF), the frequency nadir, and the quasi-steady-state (QSS) frequency deviation~\cite{BadesaTeng-4357}.
        However, these metrics are difficult to calculate because frequency dynamics are governed by a series of differential-algebraic equations (DAEs)~\cite{ZhaoWei-267}.
        Additionally, complex control process, such as the dead band that is utilized to avoid frequent PFR and to reduce the maintenance cost of governors, needs to be considered yet further complicates the computation~\cite{LiuBizzarri-4683}.
        Much effort has been devoted to simulating or approximating frequency dynamics and TABLE \ref{tab:nadir} provides a review and comparison of existing methods and ours.
        Numerical simulation tools (such as Simulink), which are commonly utilized in transient analysis, can encode complex control strategies and thus accurately evaluate frequency dynamics.
        Nonetheless, these methods are computationally heavy and intractable to compute in optimization models.
        For frequency stability-constrained optimization, existing approaches are mainly based on the system frequency response (SFR) model, which have been extended in various ways.
        The first variant is dynamic response models, which consider the frequency dynamics~\cite{ZhangZhou-3475, LiQiao-1435, AhmadiGhasemi-4685, ZhangDu-4684, LiuHu-5670, ZhangWu-5669, PaturetMarkovic-5673, ShenWu-5674, JavadiAmraee-5672, JavadiGong-5671}.
        The difference method is a straightforward method to handle the DAEs of the model \cite{JavadiAmraee-5672, JavadiGong-5671}, in which small time steps are required for accurate modeling but it incurs a heavy computational burden.
        Most works on dynamic response models assume the response time of different PFR resources to be the same and ignore the dead band, through which frequency dynamics and its nadir can be analytically derived.
        Yet, these assumptions are less practical and the derived analytical expression of nadir is highly nonlinear, for which complex approximation methods are further required to handle the intractability, such as piecewise linearization~\cite{ZhangZhou-3475, LiQiao-1435, AhmadiGhasemi-4685, ZhangDu-4684,   LiuHu-5670, ZhangWu-5669} and inner approximations~\cite{PaturetMarkovic-5673, ShenWu-5674}.
        Second, some studies simplify the dynamic response model by assuming a linear ramping for PFR, wherein the dead band can be embedded~\cite{YangXu-3622, ChuZhang-2344, YangPeng-5678, LiAi-5667, BadesaTeng-5028, Trovato-5774}.
        But such simplification ignores PFR dynamics and induces low accuracy.
        Moreover, the approach also results in nonlinear expressions for frequency nadir constraints, and consequently, sufficient conditions are adopted for tractable calculation~\cite{YangXu-3622, ChuZhang-2344, YangPeng-5678, LiAi-5667, BadesaTeng-5028, Trovato-5774}, most of which are unfortunately conservative and sacrifice the economy.
        Finally, other variants consider machine learning techniques and utilize neural networks to fit the nonlinear frequency dynamics \cite{ZhangCui-2653, SangXu-5303}.
        With proper sampling and training, such neural network-based methods can capture the dynamic characteristics to yield an accurate prediction of frequency dynamics.
        Nevertheless, few have considered the dead band and the neural network is computationally prohibitive to use in optimization models.
        In this paper, we aim to model frequency security constraints that incorporate both frequency dynamics and the dead band into UC in a tractable manner.

        \begin{table*}[!t]
            \renewcommand{\arraystretch}{1.25}
            \vspace{-0ex}
            \caption{Comparison of Main Methods for Calculating the Frequency Nadir}
            \vspace{-2ex}
            \label{tab:nadir}
            \centering
            \newcommand{\tabincell}[2]{\begin{tabular}{@{}#1@{}}#2\end{tabular}}
            \begin{tabular}{|c|c|c|c|c|c|c|}
                \hline
                \multicolumn{2}{|c|}{Method} & \tabincell{c}{Response\\dynamics} &  \tabincell{c}{Different\\response constants} & \tabincell{c}{Dead\\band} & \tabincell{c}{Encoding nadir constraints} & Reference \\
                \hline
                \multicolumn{2}{|c|}{Numerical simulation} & Yes & Yes  & Yes & N/A & MATLAB/Simulink \\
                \hline
                \multicolumn{2}{|c|}{Neural network prediction} & Yes & Yes & Yes & \tabincell{c}{Mixed-integer linearization} & \cite{ZhangCui-2653, SangXu-5303} \\
                \hline
                \multicolumn{2}{|c|}{Linear ramping model} & No & Yes & Yes & Sufficient condition & \cite{YangXu-3622, ChuZhang-2344, YangPeng-5678, LiAi-5667, BadesaTeng-5028, Trovato-5774} \\
                \hline
                \multirow{4}{*}{\tabincell{c}{Dynamic\\response\\model}} & \multirow{2}{*}{\tabincell{c}{Analytical expression}} & \multirow{2}{*}{Yes} & \multirow{2}{*}{No} & \multirow{2}{*}{No} & Piecewise linearization & \cite{ZhangZhou-3475, LiQiao-1435, AhmadiGhasemi-4685, ZhangDu-4684,   LiuHu-5670, ZhangWu-5669} \\
                \cline{6-7}
                 &  &  &  &  & Inner approximation & \cite{PaturetMarkovic-5673, ShenWu-5674} \\
                \cline{2-7}
                 & \multirow{2}{*}{\tabincell{c}{DAE constraints}} & \multirow{2}{*}{Yes} &  \multirow{2}{*}{Yes} & \multirow{2}{*}{Yes} & Difference method & \cite{JavadiAmraee-5672, JavadiGong-5671} \\
                \cline{6-7}
                 &  &  &   &  & Our method & -- \\
                \hline
            \end{tabular}
            \vspace{-4pt}
        \end{table*}

        Renewable generation is uncertain, which strongly impacts PFR capacities and frequency dynamics.
        In what follows, we consider wind power as an example~\cite{JiangWang-944} and the same approach adapts to other renewables.
        Due to wind uncertainties, the real-time available wind generation may be different from the day-ahead forecast that is used to decide the PFR reserves of wind farms.
        If the available wind generation is less than the required PFR reserves, PFR capacities from wind farms will be limited, resulting in a larger frequency deviation and threatening frequency security~\cite{GalianaBouffard-4678, VidyanandanSenroy-4364}.
        Two approaches have been studied to address this issue.
        First, the frequency stability-constrained UC model explicitly incorporates wind uncertainties using, e.g., stochastic approximation~\cite{LiQiao-1435}, chance constraints~\cite{DingZeng-1190, ZhengLiao-4620, YangXu-3622, ChuZhang-2344, ZhangShen-828}, and robust optimization~\cite{JiangBie-4478, ZhouFang-1819}.
        In this paper, we adopt distributionally robust chance constraints (DRCCs) in view of their tractability and good out-of-sample performance~\cite{li2019ambiguous, shen2021convex, shen2022wasserstein}.
        Second, we adjust PFR reserves across time, e.g., through the droop factors used in the control strategy of wind farm inverters~\cite{YuanZhang-4557}, in response to varying wind generation.
        We note that the droop factors in~\cite{YuanZhang-4557} are predefined device parameters, which vary for different frequency deviation but remain constant across time.
        In contrast, we consider droop factors as decision variables and optimize their values over time.
        Variable droop factors also help to better coordinate thermal units and wind farms in providing frequency support and enhancing the operational economy.
        Therefore, both DRCCs and variable droop factors are considered in this work to better accommodate wind uncertainties.

        \vspace{-2ex}
        \subsection{Contributions and Paper Organization}
        Considering the time-domain characteristic of frequency dynamics, Bernstein polynomials (BPs) are promising to handle the above research gaps due to their tractability for continuous-time optimization \cite{ParvaniaScaglione-544}, which has shown superiority in exploiting flexibility \cite{ZhouFang-2644, ZhouFang-4358} and modeling natural gas transmission \cite{ZhengFang-1987}.
        This paper proposes a tight, Bernstein polynomials (BPs)-based approximation method for frequency stability-constrained UC, which enables UC to incorporate more details of frequency dynamics (e.g., different response constants, the dead band and variable droop factors) while keeping the model tractability.
        Our main contributions are summarized as follows.
        \begin{enumerate}
            \item We propose a frequency stability-constrained UC model to accommodate under-frequency events, which incorporates time-domain frequency dynamics, different response constants, the dead band control, and variable droop factors.
                Thermal units and wind farms are coordinated to provide frequency support.
            \item We develop a tight, BP-based approximation to transform time-domain frequency security constraints, which involve intractable DAEs, into mixed-integer linear constraints that can be computed by off-the-shelf solvers.
            \item Through extensive case studies, we demonstrate the tightness of the BP approximation, the necessity of considering the dead band, the value of using variable droop factors, and the scalability of our approach.
        \end{enumerate}

        The remainder of this paper is organized as follows.
        Section~\ref{sec:formulation} establishes the problem formulation of frequency stability-constrained UC under wind power uncertainty.
        Section~\ref{sec:algorithm} derives the BP approximation of frequency dynamics DAEs and a tractable reformulation of the DRCCs.
        Case studies are conducted in Section~\ref{sec:results} and conclusions are drawn in Section~\ref{sec:conlcusion}.

    \section{Problem Formulation}\label{sec:formulation}
        We first introduce the general process of frequency dynamics, then propose the DAE frequency security constraints, and finally formulate a frequency stability-constrained UC model.

        \subsection{Frequency Dynamics}
\begin{figure}[!htbp]	
                \centering
                \includegraphics[width=0.45\textwidth]{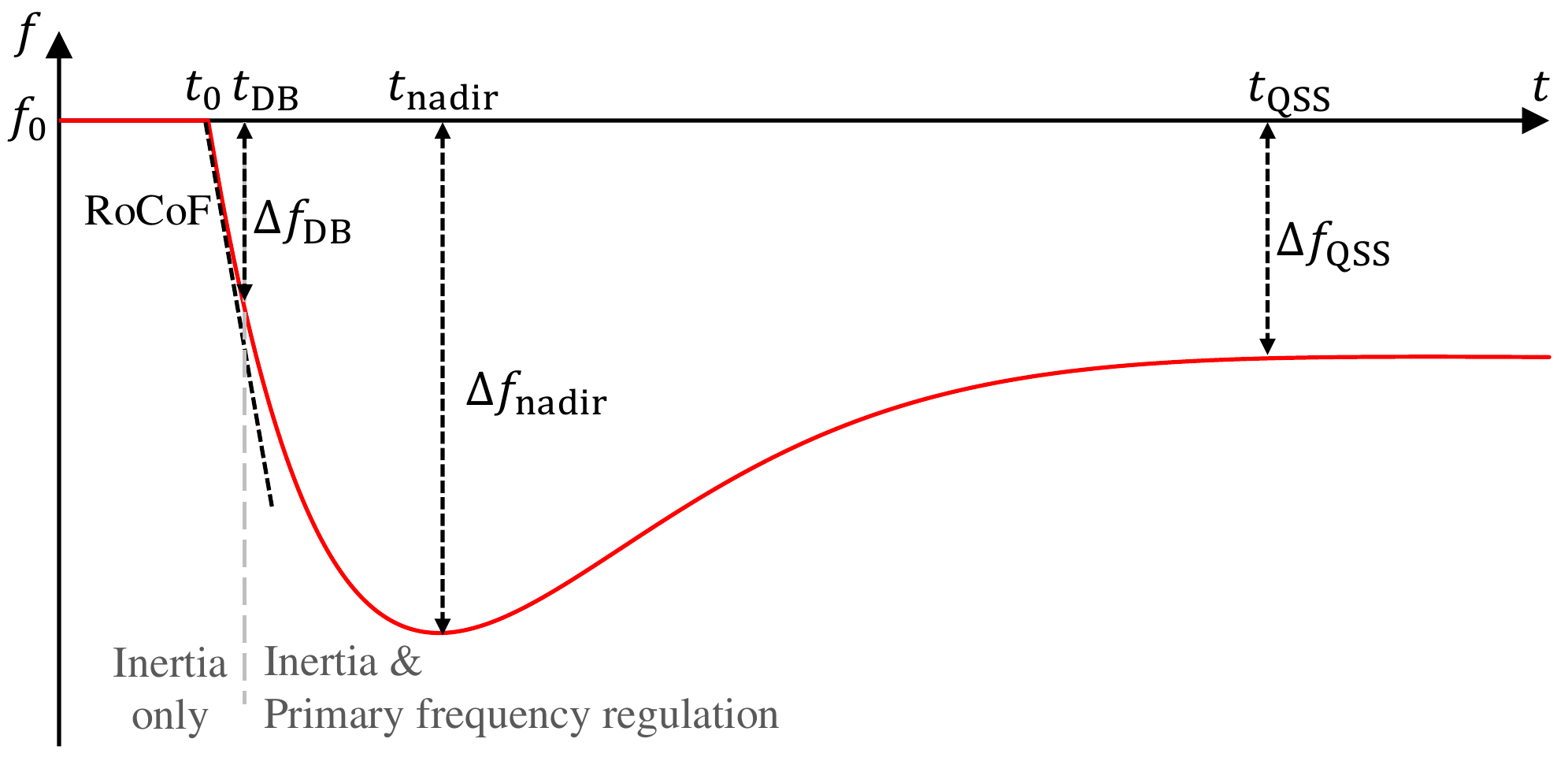}
                \caption{A general process of the frequency dynamics}
                \label{fig:dynamics}
            \end{figure}

            Fig.~\ref{fig:dynamics} depicts a general process of frequency dynamics under a sudden loss of generation at time $t_{0}$.
            At the beginning (time $t_{0}$--$t_{\textrm{DB}}$), before the frequency deviation exceeds the dead band $\Delta f_{\textrm{DB}}$, the governors do not work and inertia plays the whole role in mitigating the frequency drop.
            Generally, the maximum RoCoF occurs at $t_{0}$.
            As the frequency deviation exceeds $\Delta f_{\textrm{DB}}$, the governors start working, and PFR becomes significant (time $t_{\textrm{DB}}$--$t_{\textrm{QSS}}$), the frequency eventually reaches a QSS with deviation $\Delta f_{\textrm{QSS}}$.
            During the whole dynamics, frequency first drops and is then lifted, yielding a nadir with deviation $\Delta f_{\textrm{nadir}}$ at $t_{\textrm{nadir}}$.

        \subsection{DAE Frequency Security Constraints}
\begin{figure}[!htbp]	
            \centering
            \includegraphics[width=0.45\textwidth]{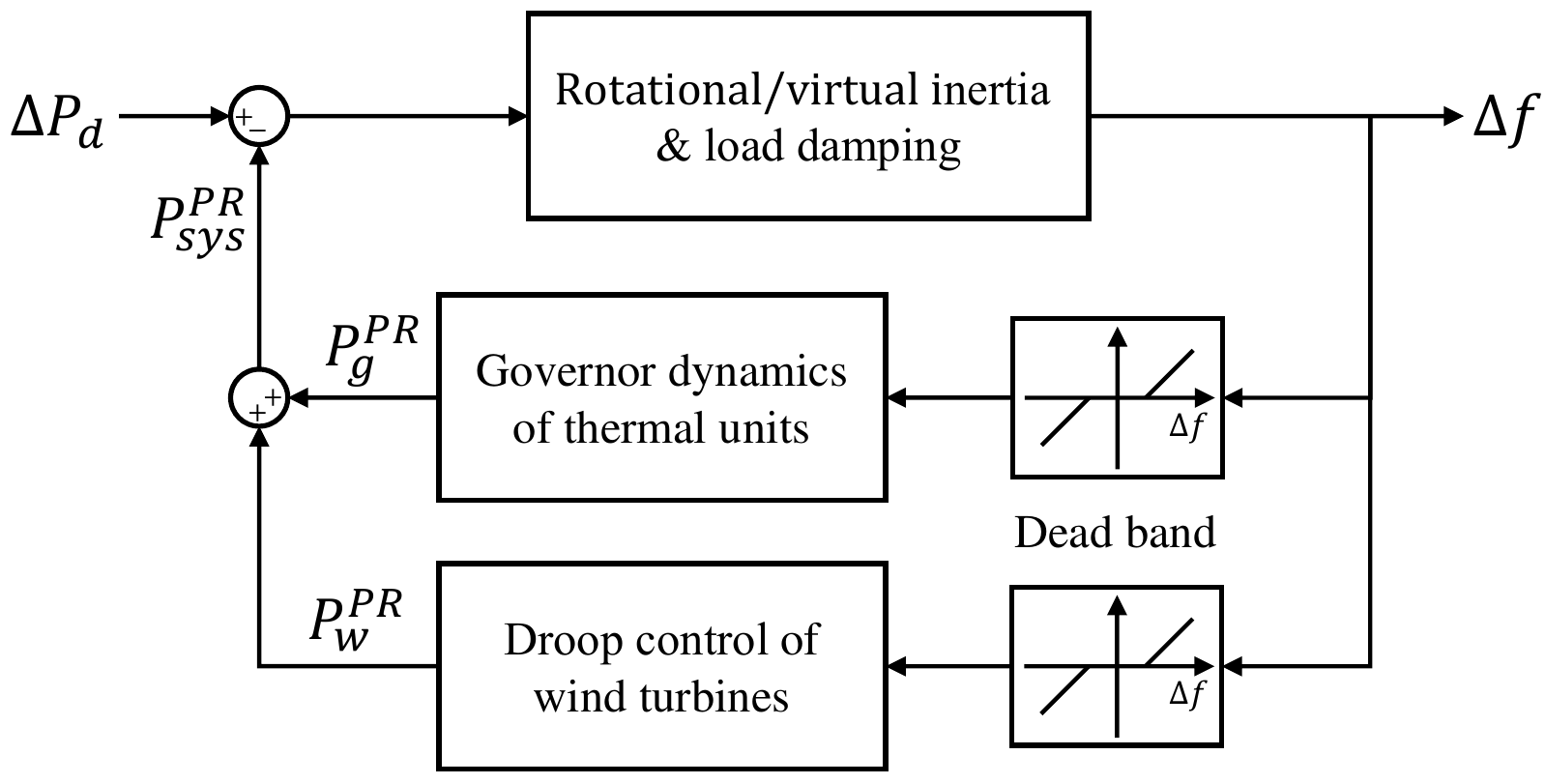}
            \caption{SFR model considering the dead band}
            \label{fig:control}
        \end{figure}

        We depict the SFR model~\cite{ZhaoWei-267} of frequency dynamics in Fig.~\ref{fig:control}.
        We note that the paper uses the center of inertia (COI) assumption, which calculates average frequency and ignores local frequency dynamics.
        In this model, $\Delta P_{d}$ is the power imbalance, $\Delta f(t)=f_{0}-f(t)$ is the frequency deviation from the nominal frequency $f_{0}$ at time instant $t$, $P^{PR}_{sys}$ is the total PFR power in the power system, and $P^{PR}_{g}$ and $P^{PR}_{w}$ are the PFR power from thermal units and wind farms, respectively.
        To avoid frequent PFR, the governors of thermal units and inverters of wind farms will not respond to a non-zero $\Delta f(t)$ unless its magnitude exceeds the dead band $\Delta f_{\text{DB}}$.

        This divides the frequency dynamics into two stages.
        In the first stage, frequency deviation lies within the dead band and thus there is no PFR.
        As in Fig.~\ref{fig:control}, system frequency dynamics follow the governing equations
        \begin{subequations}\label{eq:dynamicDB}
            \begin{gather}
                    2H_{sys}\frac{\textrm{d}\Delta f(t)}{\textrm{d}t}+k_{D}P_{d}\Delta f(t)=\Delta P_{d},~t\in[0,t_{DB}]\label{eq:dynamicDB-system}\\
                    \Delta f(t)|_{t=0}=0,~\Delta f(t)|_{t=t_{DB}}=\Delta f_{DB},\label{eq:dynamicDB-initial}
            \end{gather}
        \end{subequations}
        where $H_{sys}$ is the total inertia in the power system, including rotational inertia $H_{g}$ from thermal units and virtual inertia $H_{w}$ from wind farms, $k_{D}$ is the load damping rate, and $P_{d}$ is the power load.
        We note that $H_{w}$ is implemented through the virtual inertia control of wind turbines \cite{AraniEl-Saadany-5263, FangLi-5264}, which is adjustable but is treated as an input parameter in this paper.
        Constraint~\eqref{eq:dynamicDB-initial} gives the initial and final conditions of the ordinary differential equation~\eqref{eq:dynamicDB-system}.
        Note that $t_{DB}$ in~\eqref{eq:dynamicDB} is not a given parameter, but a variable that depends on frequency dynamics and the dead band $\Delta f_{DB}$.
        In the second stage, frequency deviation exceeds $\Delta f_{DB}$ and  PFR starts working.
        Accordingly, the system frequency dynamics follow the governing equations
        \begin{subequations}\label{eq:dynamic}
            \begin{gather}
                    2H_{sys}\frac{\textrm{d}\Delta f(t)}{\textrm{d}t}+k_{D}P_{d}\Delta f(t)=\Delta P_{d}-P^{PR}_{sys}(t),~t\geq t_{DB}\label{eq:dynamic-system}\\
                    \Delta f(t)|_{t=t_{DB}}=\Delta f_{DB}\label{eq:dynamic-initial}
            \end{gather}
        \end{subequations}
        \vspace{-3ex}
        \begin{subequations}\label{eq:PR}
            \begin{gather}
                    P^{PR}_{sys}(t)=\sum_{i}\left[P^{PR}_{g,i}(t)+P^{PR}_{w,i}(t)\right]\label{eq:PR-total}\\
                    T_{g,i}\frac{\textrm{d}P^{PR}_{g,i}(t)}{\textrm{d}t}+P^{PR}_{g,i}(t)=G_{g,i}I_{g,i}\left[\Delta f(t)-\Delta f_{DB}\right]\label{eq:PR-TU}\\
                    P^{PR}_{g,i}(t)|_{t=t_{DB}}=0\label{eq:PR-TU-initial}\\
                    P^{PR}_{w,i}(t)=G_{w,i}\left[\Delta f(t)-\Delta f_{DB}\right],\label{eq:PR-WF}
            \end{gather}
        \end{subequations}
        where $i$ is the bus index, $T_{g}$ is the response constant of governors, and $G_{g}$ is the droop factor of thermal units.
        The response constant of wind power is $T_{w}\approx0$ because it is controlled by inverters.
        $G_{w}$ is the variable droop factor of wind farms, which can be adjusted to adapt to wind uncertainties and to enhance the coordination of thermal units and wind farms.
        Constraint~\eqref{eq:dynamic-initial} gives the initial condition of the ordinary differential equation~\eqref{eq:dynamic-system}.
        Constraint~\eqref{eq:PR-total} implies that the total PFR power $P^{PR}_{sys}(t)$ comes from all power sources, including thermal units and wind farms.
        According to the response characteristic of the governors, \eqref{eq:PR-TU} describes the dynamics of $P^{PR}_{g}$, where
        $I_{g}$ indicates the online/offline status of thermal units ($I_g = 1$ if online and $I_g = 0$ if offline).
        Constraint~\eqref{eq:PR-TU-initial} provides the initial condition of the ordinary differential equation~\eqref{eq:PR-TU}.
        Finally, similar to~\eqref{eq:PR-TU}, \eqref{eq:PR-WF} describes the dynamics of the wind farm PFR power $P^{PR}_{w}$.
        Note that we consider non-step dead band control in~\eqref{eq:PR-TU} and \eqref{eq:PR-WF}, but the step implementation can also be tractably incorporated with slight adaption.

        To protect frequency security, we restrict the RoCoF, nadir, and quasi-steady-state deviation as
        \begin{subequations}\label{eq:security}
            \begin{gather}
                    \frac{\textrm{d}\Delta f(t)}{\textrm{d}t}\leq \overline{\dot{f}}\label{eq:security-RoCoF}\\
                    \Delta f(t)|_{\textrm{nadir}}\leq \overline{\Delta f}\label{eq:security-nadir}\\
                    \Delta f(t)|_{\textrm{QSS}}\leq \overline{\Delta f_{err}},\label{eq:security-QSS}
            \end{gather}
        \end{subequations}
        where $\overline{\dot{f}}$ is the maximum admitted RoCoF, and $\overline{\Delta f}$ and $\overline{\Delta f_{err}}$ are the maximum admitted frequency deviation at the nadir and QSS, respectively.
        Note that without loss of generality, only under-frequency events are considered.

        \subsection{Frequency Stability-Constrained Chance-Constrained Unit Commitment}
            We formulate the following frequency stability-constrained UC model with DAEs and DRCCs.
            \begin{align}
                \min & \sum_{\tau}\sum_{i}\left(c_{su,i}U_{g,i,\tau}+c_{sd,i}D_{g,i,\tau}+F_{g,i,\tau}+c^{PR}_{g}R^{PR}_{g,i,\tau}+c^{PR}_{w}R^{PR}_{w,i,\tau}\right) \label{eq:objective}
            \end{align}
            \vspace{-3ex}
            \begin{equation}\label{eq:frequency-dynamics}
                \textrm{s.t.}~(\ref{eq:dynamicDB})-(\ref{eq:security}) \nonumber
            \end{equation}
            \begin{align}
            & \ \left\{
                    \begin{aligned} &F_{g,i,\tau}=c_{on,i}I_{g,i,\tau}+\sum_{k}\lambda_{g,i,k}p_{g,i,k,\tau}\\          &P_{g,i,\tau}=\underline{P_{g,i}}I_{g,i,\tau}+\sum_{k}p_{g,i,k,\tau}\\
                    &0\leq p_{g,i,k,\tau}\leq (\overline{P_{g,i}}-\underline{P_{g,i}})/N_{pw}
                    \end{aligned}
                \right. \label{eq:fuel cost}
            \end{align}
            \vspace{-1ex}
            \begin{subequations}\label{eq:first-stage}
                \begin{align}
                    & \ U_{g,i,\tau}+D_{g,i,\tau}\leq1\\
                    & \ U_{g,i,\tau}-D_{g,i,\tau}=I_{g,i,(\tau+1)}-I_{g,i,\tau}\\
                    & \ \left\{
                        \begin{aligned}
                            &\sum_{k=\tau+1}^{t+T_{on,i}}I_{g,i,k}\geq T_{on,i}U_{g,i,\tau}\\
                            &\sum_{k=\tau+1}^{t+T_{off,i}}(1-I_{g,i,k})\geq T_{off,i}D_{g,i,\tau}
                        \end{aligned}
                    \right.
                \end{align}
            \end{subequations}
            \vspace{-2ex}
            \begin{subequations}\label{eq:frequency}
                \begin{align}
                & \ H_{sys,\tau}=\sum_{i}\left(H_{g,i}I_{g,i,\tau}+H_{w,i}\right)\label{eq:frequency-inertia}\\
                & \ \underline{G_{w,i}}\leq G_{w,i,\tau}\leq\overline{G_{w,i}}\\
                & \ \left\{
                        \begin{aligned}
                            &P^{PR}_{g,i,\tau}(t)\leq R^{PR}_{g,i,\tau}\\
                            &P^{PR}_{w,i,\tau}(t)\leq R^{PR}_{w,i,\tau}
                        \end{aligned}
                    \right.\label{eq:frequency-reserve}
                \end{align}
            \end{subequations}
            \vspace{-1ex}
            \begin{subequations}\label{eq:unit}
                \begin{align}
                & \ \underline{P_{g,i}}I_{g,i,\tau}\leq P_{g,i,\tau}\leq\overline{P_{g,i}}I_{g,i,\tau}-R^{PR}_{g,i,\tau}\\
                & \ \left\{
                        \begin{aligned}
                            &P_{g,i,(\tau+1)}-P_{g,i,\tau}\leq r_{u,i}I_{g,i,\tau}+r_{su,i}U_{g,i,\tau}\\
                            &P_{g,i,\tau}-P_{g,i,(\tau+1)}\leq r_{d,i}I_{g,i,(\tau+1)}+r_{sd,i}D_{g,i,\tau}
                        \end{aligned}
                    \right.
                \end{align}
            \end{subequations}
            \begin{align}\label{eq:power balance}
            & \ \sum_{i}(P_{g,i,\tau}+P_{w,i,\tau})=\sum_{i}P_{d,i,\tau}
            \end{align}
            \begin{equation}\label{eq:power flow}
                -\overline{P_{l,j}}\leq\sum_{i}S_{ji}(P_{g,i,\tau}+P_{w,i,\tau}-P_{d,i,\tau})\leq\overline{P_{l,j}}
            \end{equation}
            \vspace{-1ex}
            \begin{align}\label{eq:DRCC}
            & \ \inf_{{\mathbb{P}}\in{\mathcal{D}}}{\mathbb{P}}\left\{P_{w,i,\tau}+R^{PR}_{w,i,\tau}\leq P^{A}_{w,i,\tau}\right\}\geq1-\epsilon,
            \end{align}
            where $\tau$ is the index for time periods.
            The objective~\eqref{eq:objective} consists of start-up/shut-down cost, PFR reserve cost, and fuel cost.
            Binary decision variables $U_{g}$ and $D_{g}$ represent the start-up and shut-down actions of thermal units, respectively, and $c_{su}$ and $c_{sd}$ are the corresponding costs.
            Continuous decision variables $R^{PR}_{g}$ and $R^{PR}_{w}$ represent the PFR reserves of thermal units and wind farms, respectively, and $c^{PR}_{g}$ and $c^{PR}_{w}$ are the corresponding unit costs.

            Constraints~\eqref{eq:fuel cost} formulate the fuel cost $F_{g}$ of thermal units as a piecewise linear function of the amount $P_{g}$ of power generation~\cite{ZhouFang-3859}.
            Specifically, $p_{g}$ is the generation segment, $N_{pw}$ is the piecewise number, $\lambda_{g,k}$ is the cost incremental coefficient of the $k$th segment, and $\overline{P_{g}}$ and $\underline{P_{g}}$ are the maximum and minimum power generation of thermal units, respectively.
            Constraints~\eqref{eq:first-stage} formulate the UC restrictions, where $T_{\text{on}}$ and $T_{\text{off}}$ are minimum up and down time, respectively.
            Constraints~(\ref{eq:first-stage}a)--(\ref{eq:first-stage}b) formulate the logic among decision variables $U_{g}$, $D_{g}$, and $I_{g}$, and constraints~(\ref{eq:first-stage}c) formulate the minimum up/down time restrictions.
            Constraints~\eqref{eq:frequency} formulate the frequency security, where $\overline{G_{w,i}}$ and $\underline{G_{w,i}}$ are the maximum and minimum values of the inverter droop factor, respectively.
            In particular, \eqref{eq:frequency-inertia} calculates the total inertia of the power system, and~\eqref{eq:frequency-reserve} ensures that the PFR power does not exceed the PFR reserve.
            We note that the energy of inertial response and droop control comes from PFR reserve and the kinetic energy of rotors, respectively, and hence, the inertial response is not limited by \eqref{eq:frequency-reserve}.
            Constraints~\eqref{eq:unit} describe the ramping restrictions, where $r_{u}/r_{d}$ is the upward/downward ramping limit and $r_{su}/r_{sd}$ is the maximum power generation at start-up/shut-down.
            Constraints~\eqref{eq:power balance} describe the system power balance, where $P_{d}$ is the power load.
            Constraints~\eqref{eq:power flow} describe the line capacity limits, where $j$ is the line index, $\overline{P_{l}}$ is the line capacity limit, and $S_{ji}$ is the power transfer distribution factor based on DC power flow.

            DRCCs~\eqref{eq:DRCC} ensure the sufficiency of wind power with high probability, where $P^{A}_{w}$ and $P_{w}$ represent the available and integrated power of wind farms, respectively, where $1-\epsilon$ is a pre-specified risk threshold. We adopt a Wasserstein ball $\mathcal{D}$ to describe the wind power ambiguity with
            \begin{subequations}\label{eq:ambiguity}
                \begin{gather}
                    \mathcal{D}=\left\{
                        {\mathbb{P}}\left|
                            d_{W}({\mathbb{P}},{\mathbb{P}}_{N})\leq r
                        \right.
                    \right\}\label{eq:ambiguity-set}\\
                    d_{W}({\mathbb{P}},{\mathbb{P}}_{N})=\inf_{(\omega,\omega_{N})\sim({\mathbb{P}},{\mathbb{P}}_{N})}{\mathbb{E}}\left[
                        \parallel\omega-\omega_{N}\parallel
                    \right],\label{eq:ambiguity-distance}
                \end{gather}
            \end{subequations}
            where ${\mathbb{P}}$ represents an ambiguous distribution of the wind power, ${\mathbb{P}}_{N}=N(\mu_{i,\tau},\sigma^{2}_{i,\tau})$ is the Gaussian distribution with the mean $\mu$ and standard deviation $\sigma$, and $r$ is the radius of $\mathcal{D}$.
            Here, $\mu$ equals to the forecast of wind power.

    \section{Solution Algorithm}\label{sec:algorithm}
    The proposed formulation~\eqref{eq:objective}--\eqref{eq:DRCC} is a mixed-integer nonlinear program due to the DAEs and DRCCs.
    In this section, we recast them as tractable and computable forms.
    We first derive an inner approximation for DAEs, and then reformulate it as well as the DRCCs as linear constraints.

        \subsection{Transformation of Frequency Security Metrics}
        To handle DAE frequency security constraints, we revisit the three metrics in~\eqref{eq:security}.
        As analyzed in Section \ref{sec:formulation}.A, since the maximum RoCoF occurs at $t=0$, constraint~\eqref{eq:security-RoCoF} is equivalent to
            \begin{equation}\label{eq:security-RoCoF-SST}
                2\overline{\dot{f}}H_{sys,\tau}\geq\Delta P_{d,\tau}.
            \end{equation}

            In addition, at the QSS, since $\Delta f(t)$ reaches a steady value, $\Delta f(t)|_{QSS}$ and $P^{PR}_{(\cdot)}(t)$ are unchanging.
            In other words, both $\textrm{d}\Delta f(t)/\textrm{d}t$ and $\textrm{d}P^{PR}_{(\cdot)}(t)/\textrm{d}t$ equal zero.
            It follows from~\eqref{eq:dynamic}--\eqref{eq:PR} that~\eqref{eq:security-QSS} is equivalent to
            \begin{subequations}\label{eq:security-QSS-SST}
                \begin{gather}
                    k_{D}P_{d}\overline{\Delta f_{err}}+G_{sys,\tau}\left(\overline{\Delta f_{err}}-\Delta f_{DB}\right)\geq\Delta P_{d,\tau}\\
                    G_{sys,\tau}=\sum_{i}\left(G_{g,i}I_{g,i,\tau}+G_{w,i}\right),
                \end{gather}
            \end{subequations}
            where $G_{sys}$ represents the sum of droop factors.

            Unlike RoCoF and QSS, $\Delta f(t)|_{\text{nadir}}$ is far more challenging to derive.
            Fortunately, the frequency nadir usually occurs beyond the dead band and so we can focus on the DAEs in~\eqref{eq:dynamic}--\eqref{eq:PR}.
            The next subsection derives an inner approximation using BPs.

        \subsection{DAE Approximation}
            We first introduce an approximation framework using BPs and then derive an inner approximation of the DAEs for computation.

            \subsubsection{BP approximation framework}~

            We restrict the functional form of the variables in the DAEs~\eqref{eq:dynamic}--\eqref{eq:PR} using BP splines~\cite{ParvaniaScaglione-544}.
            Specifically, if we restrict the functional form of a continuous-time function $F(t)$ to be cubic BPs, then
            \begin{equation}\label{eq:spline}
                F(t) = \sum_{k=0}^{3}F^{B,k}B_{3,k}(t)=({{F}}^{{{B}}})^{\textrm{T}}{{B}}_{{3}}(t),~t\in[0,1]
            \end{equation}
            where $B_{3,k}(t) := \begin{pmatrix} 3\\k \end{pmatrix} t^{k}(1-t)^{3-k}$ for all $t\in[0,1]$ and $k=0,1,2,3$ is a cubic BP, and ${{F}}^{{{B}}} := [F^{B,0}, \ldots, F^{B,3}]^{\top}$ are the spline coefficients. This automatically guarantees the differentiability of $F(t)$ and allows us to compute its derivatives and integrals exactly or approximately. In particular,
            \begin{subequations}\label{eq:SST}
                \begin{gather}
                    \frac{\textrm{d}F(t)}{\textrm{d}t}=\frac{\textrm{d}({{F^{B}}})^{\textrm{T}}{{B_{3}}}(t)}{\textrm{d}t}=({W}{{F}}^{{{B}}})^{\textrm{T}}{{B}}_{{2}}(t)\label{eq:SST-derivative}\\
                    \int_{0}^{1}F(t)\textrm{d}t={1}^{\textrm{T}}{{F}}^{{{B}}}/4\label{eq:SST-integral-01}\\
                    \int_{0}^{t}F(t)\textrm{d}t=\int_{0}^{t}({{F^{B}}})^{\textrm{T}}{{B_{3}}}(t)\textrm{d}t \approx ({{F^{B}}})^{\textrm{T}}{{LB_{3}}}(t),\label{eq:SST-integral-0t}\\
                    F(t)=0 \ \Leftrightarrow \ ({{F}}^{{{B}}})^{\textrm{T}}{{B}}_{{3}}(t)=0 \ \Leftrightarrow \ {{F}}^{{{B}}}=0\label{eq:SST-equality}\\
                    F(t)\leq0 \ \Leftrightarrow \ ({{F}}^{{{B}}})^{\textrm{T}}{{B}}_{{3}}(t)\leq0 \ \Leftarrow \ {{JF}}^{{{B}}}\leq0\label{eq:SST-inequality}
                \end{gather}
            \end{subequations}
            where $B_{2,k}(t)$ represents quadratic BPs, and $W$~\cite{ParvaniaScaglione-544}, $L$~\cite{ZhengFang-1987}, and $J$~\cite{ZhouFang-2644} are matrices of known entries.
            It can be seen that through BP approximation, all the derivatives, integrals, equality and inequality constraints of the continuous-time functions can be reformulated as their linear counterparts.
            We note that, as compared to other spline methods, BPs have some important properties: i) convex--hull property that enables the inequality transformation \eqref{eq:SST-inequality}, ii) subdivision property that strengthens the inequality transformation \eqref{eq:SST-inequality}, and iii) operation matrices that enable the integral transformation \eqref{eq:SST-integral-0t}.
            The interested readers can refer to \cite{ParvaniaScaglione-544, ZhengFang-1987, ZhouFang-4358, ZhouFang-2644} for more details.

            \subsubsection{DAE approximation}~

            We apply the BP approximation framework to derive an inner approximation of the DAEs~\eqref{eq:dynamic}--\eqref{eq:PR} and constraint~\eqref{eq:security-nadir}.

\begin{figure}[!htbp]	
                \centering
                \includegraphics[width=0.45\textwidth]{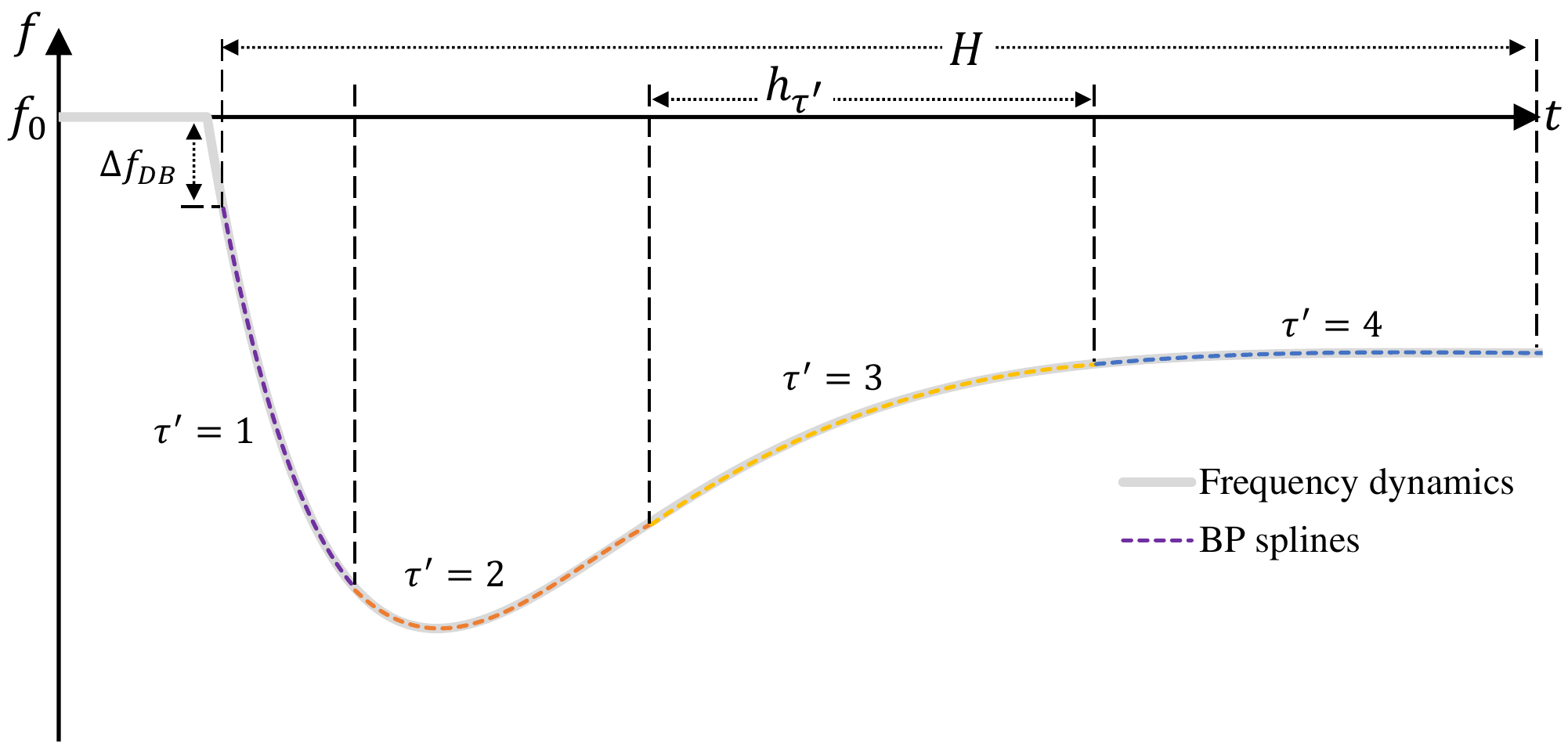}
                \caption{BP spline-based frequency dynamics}
                \label{fig:approximation}
            \end{figure}

            First, frequency dynamics are piecewise approximated using BP splines. Fig.~\ref{fig:approximation} depicts an illustrative example.
            We divide the horizon of frequency dynamics, $H$, into several segments, each with a time length $h_{\tau'}$, where $\tau'$ is the index.
            For each segment, we normalize the time range as $[0, 1]$.
            Using BP splines, we approximate $\Delta f(t)$ in each segment $\tau'$ as
            \begin{equation}\label{eq:spline2}
                \Delta f_{\tau'}(t)=({{\Delta f_{\tau'}^{B}}})^{\textrm{T}}{{B_{3}}}(t),~t\in[0,1]
            \end{equation}
            where $\Delta f^{B}$ consists of the spline coefficients.
            PFR dynamics $P^{PR}_{g}(t)$ and $P^{PR}_{w}(t)$ are similarly handled and $P^{PR,B}_{g}$ and $P^{PR,B}_{w}$ are the corresponding spline coefficients.

            Second, we integrate both sides of (\ref{eq:dynamic}a) from 0 to $t$ to yield
            \begin{equation}\label{eq:dynamic2}
                \int_{0}^{t}\left[\frac{2H_{sys}}{h_{\tau'}}\frac{\textrm{d}\Delta f(t)}{\textrm{d}t}+k_{D}P_{d}\Delta f(t)\right]\textrm{d}t=\int_{0}^{t}\left[\Delta P_{d}-\Delta P^{PR}_{sys}(t)\right]\textrm{d}t,
            \end{equation}
            which by~\eqref{eq:SST} is equivalent to
            \begin{subequations}\label{eq:dynamic-SST}
                \begin{gather}
                    \frac{2H_{sys}}{h_{\tau'}}\left({{\Delta f^{B}_{\tau'}}}-{{\Delta f_{\tau',ini}^{B}}}\right)+k_{D}P_{d}{{L}}^{\textrm{T}}{{\Delta f^{B}_{\tau'}}}={{L}}^{\textrm{T}}\left({{\Delta P_{d}^{B}}}-{{P^{PR,B}_{sys,\tau'}}}\right)\\
                    \Delta f_{\tau',ini}^{B,k}|_{\tau'=1}=\Delta f_{DB},~\Delta f_{\tau',ini}^{B,k}|_{\tau'>1}=\Delta f_{\tau'-1}^{B,3},
                \end{gather}
            \end{subequations}
            where ${\Delta f_{\tau',ini}^{B}}$ is the initial value of ${\Delta f_{\tau'}^{B}}$ in period $\tau'$ and~(\ref{eq:dynamic-SST}b) designates the initial condition for each period.

            Similarly, \eqref{eq:PR} can be recast as
            \begin{subequations}\label{eq:PR-SST}
                \begin{gather}{{P^{PR,B}_{sys,\tau'}}}=\sum_{i}\left({{P^{PR,B}_{g,i,\tau'}}}+{{P^{PR,B}_{w,i,\tau'}}}\right)\\
                    \frac{T_{g,i}}{h_{\tau'}}\left({{P_{g,i,\tau'}^{PR,B}}}-{{P_{g,i,\tau',ini}^{PR,B}}}\right)+{{L}}^{\textrm{T}}{{P_{g,i,\tau'}^{PR,B}}}=G_{g,i}I_{g,i}{{L}}^{\textrm{T}}\left({{\Delta f^{B}_{\tau'}}}-{{\Delta f^{B}_{DB}}}\right)\\
                    P_{g,i,\tau',ini}^{PR,B,k}|_{\tau'=1}=0,~P_{g,i,\tau',ini}^{PR,B,k}|_{\tau'>1}=P_{g,i,\tau'-1}^{PR,B,3}\\
                    {{P_{w,i,\tau'}^{PR,B}}}=G_{w,i}\left({{\Delta f^{B}_{\tau'}}}-{{\Delta f^{B}_{DB}}}\right),
                \end{gather}
            \end{subequations}
            where ${P^{PR,B}_{w,i,\tau',ini}}$ is the initial value of ${P^{PR,B}_{w,i,\tau'}}$ in period $\tau'$.

            Finally, by~\eqref{eq:SST-inequality}, constraints~\eqref{eq:security-nadir} and~\eqref{eq:frequency-reserve} are respectively implied by
            \begin{equation}\label{eq:security-nadir-SST}
                {{J\Delta f^{B}_{\tau'}}}\leq \overline{\Delta f}
            \end{equation}
            \begin{equation}\label{eq:reserve-SST}
            \text{and} \quad
                \begin{aligned}
                    \left\{
                        \begin{aligned}
                            &{{JP^{PR,B}_{g,i,\tau,\tau'}}}\leq R^{PR}_{g,i,\tau}\\
                            &{{JP^{PR,B}_{w,i,\tau,\tau'}}}\leq R^{PR}_{w,i,\tau}.
                        \end{aligned}
                    \right.
                \end{aligned}
            \end{equation}

            Consequently, we inner approximate the DAE frequency security constraints by general algebraic constraints~\eqref{eq:security-RoCoF-SST}, \eqref{eq:security-QSS-SST}, and~\eqref{eq:dynamic-SST}--\eqref{eq:reserve-SST}.

    \vspace{-2ex}
        \subsection{Linearization}
            We reformulate the bilinear terms $H_{sys}{{\Delta f^{B}_{\tau'}}}$ in (\ref{eq:dynamic-SST}a), $I_{g,i}{{\Delta f^{B}_{\tau'}}}$ in (\ref{eq:PR-SST}b), and $G_{w,i}{{\Delta f^{B}_{\tau'}}}$ in (\ref{eq:PR-SST}d) as linear forms for tractable computation. In what follows, we linearize $G_{w,i}{{\Delta f^{B}_{\tau'}}}$ as an example, and the linearization of other bilinear terms can be conducted similarly.
            \subsubsection{Binary expansion}
            Assuming that the droop factor $G_{w}$ can only be adjusted discretely with a minimum stepsize $\Delta G_{w}$~\cite{ZhouFang-4358}, we encode $G_w$ using binary decision variables $\omega_k$ as
            \begin{equation}\label{eq:binary expansion}
                G_{w}=\sum_{k=0}^{N_{BE}-1}\omega_{k}2^{k}\Delta G_{w},~\omega_{k}\in\{0,1\},
            \end{equation}
            where $N_{BE}$ is a pre-specified number for $\omega_{k}$ to adjust the granularity of the binary expansion.

            \subsubsection{Big-M method}
            With the above binary expansion, the bilinear term $G_{w}{{\Delta f^{B}_{\tau'}}}$ can be rewritten as ${{\Delta f^{B}_{\tau'}}}\sum_{k}\omega_{k}2^{k}\Delta G_{w}$.
            Then, through the big-M method, the new bilinear terms ${{\alpha_{k,\tau'}}}=\omega_{k}{{\Delta f^{B}_{\tau'}}}$ can be linearized as
            \begin{subequations}\label{eq:bigM}
                \begin{gather}
                    -M\omega_{k}\leq{{\alpha_{k,\tau'}}}\leq M\omega_{k}\\
                    -M(1-\omega_{k})+{{\Delta f^{B}_{\tau'}}}\leq{{\alpha_{k,\tau'}}}\leq{{\Delta f^{B}_{\tau'}}}+M(1-\omega_{k}).
                \end{gather}
            \end{subequations}

            \subsubsection{DRCC Reformulation}
            We follow~\cite{shen2021convex} to recast the DRCC~\eqref{eq:DRCC} as
            \begin{equation}\label{eq:DRCC2}
                P_{w,i,\tau}+R^{PR}_{w,i,\tau}\leq \mu_{i,\tau}-c_{p}\sigma_{i,\tau}
            \end{equation}
            where $c_{p}$ is a parameter that only depends on $\epsilon$ and $r$ and can be computed beforehand.

            We have recast the proposed frequency stability-constrained UC formulation~\eqref{eq:objective}--\eqref{eq:DRCC} as a mixed-integer linear program, which can be directly computed by off-the-shelf solvers.

    \section{Case Study}\label{sec:results}
        We first demonstrate the tightness of the BP approximation. Then, using the IEEE 6-bus system, we analyze frequency security, influence of the dead band, variable droop factors, and parameter settings of the DRCCs. Finally, we demonstrate the scalability of our approach using an IEEE 118-bus system.

        \vspace{-2ex}
        \subsection{Tightness of Frequency Dynamics}
\begin{figure}[!t]	
                    \centering
                    \includegraphics[width=0.35\textwidth]{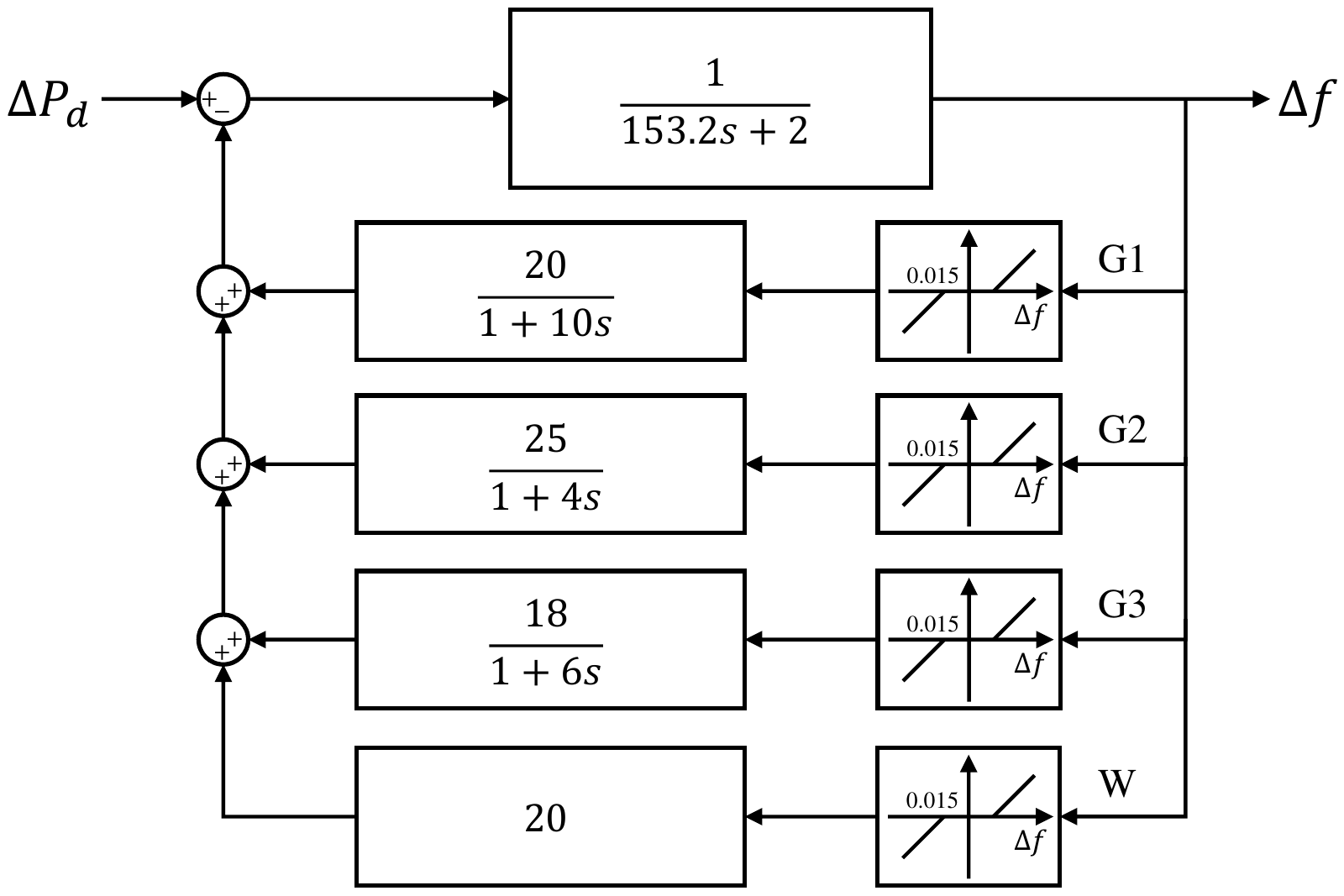}
                    \caption{SFR model with three thermal units and a wind farm}	
                    \label{fig:SFR}
            \end{figure}

            We demonstrate the tightness of the BP approximation~\eqref{eq:dynamic-SST}--\eqref{eq:PR-SST} in evaluating the nadir.
            We consider three thermal units and a wind farm and their parameters are the same as in the 6-bus system in Section~\ref{sec:6-bus} (see Table~\ref{tab:parameter}), except that for now we fix the droop factor of the wind farm at 20 MW/Hz. We illustrate the corresponding SFR model in Fig.~\ref{fig:SFR}, which is consistent with the frequency security constraints \eqref{eq:dynamicDB}--\eqref{eq:PR}.
            \begin{figure}[!t]	
                \centering
                \subfigure[Different lengths of horizon]{\includegraphics[width=0.45\textwidth]{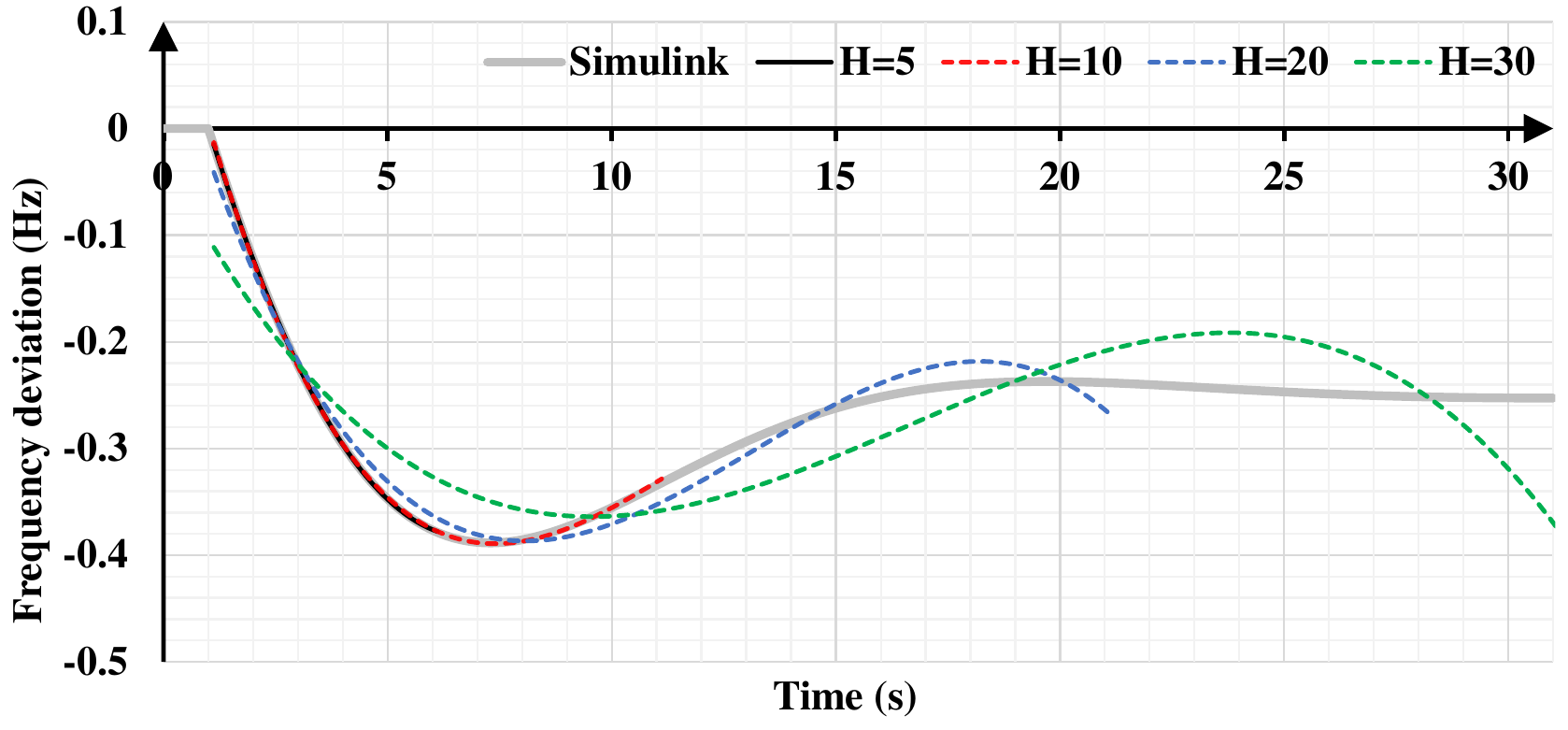}}\\
                \vspace{-1ex}
                \subfigure[Different segment number]{\includegraphics[width=0.45\textwidth]{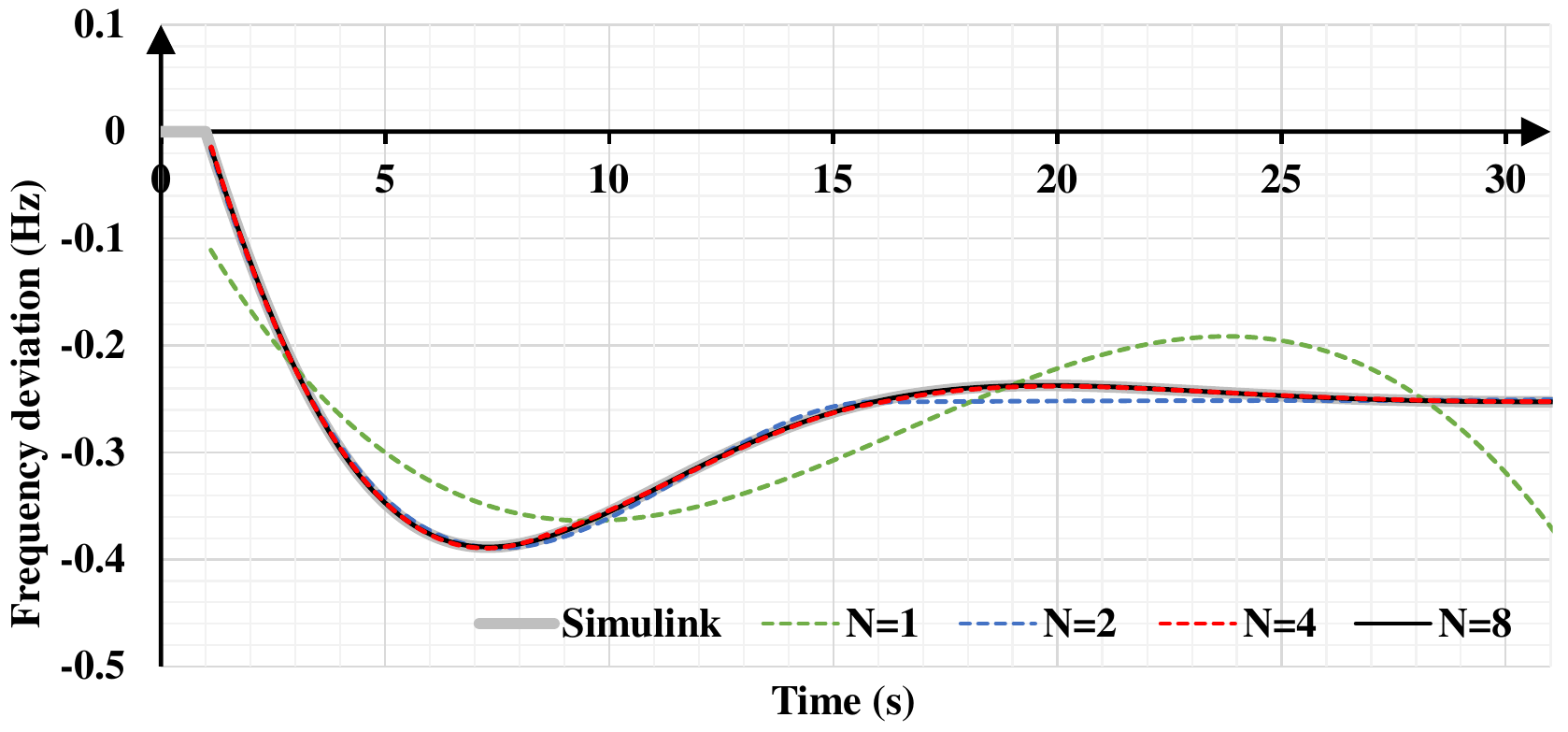}}
                \vspace{-1ex}
                \caption{Frequency dynamics from BP approximation vs. from Simulink}	
                \label{fig:accuracy-even}
            \end{figure}
            \begin{table}[!t]
                \renewcommand{\arraystretch}{1.1}
                \vspace{-2ex}
                \caption{Nadir Values and Their Relative Errors under Different $H$}
                \vspace{-2ex}
                \label{tab:accuracy-horizon}
                \centering
                \newcommand{\tabincell}[2]{\begin{tabular}{@{}#1@{}}#2\end{tabular}}
                \begin{tabular}{cccccc}
                    \toprule
                     & \multirow{2}{*}{Simulink} & \multicolumn{4}{c}{BP approximation}\\
                    \cmidrule{3-6}
                     &  & $H$=5s & $H$=10s & $H$=20s & $H$=30s\\
                    \midrule
                    Nadir value (Hz) & \hspace{-1ex}$-0.3884$\hspace{-1ex} & \hspace{-1ex}$-0.3778$\hspace{-1ex} & \hspace{-1ex}$-0.3892$\hspace{-1ex} & \hspace{-1ex}$-0.3866$\hspace{-1ex} & \hspace{-1ex}$-0.3759$\hspace{-1ex}\\
                    Relative error & N/A & $2.73\%$ & $0.20\%$ & $0.46\%$ & $3.21\%$\\
                    \bottomrule
                \end{tabular}
                \vspace{-0ex}
            \end{table}
            \begin{table}[!t]
                \renewcommand{\arraystretch}{1.1}
                \vspace{-2ex}
                \caption{Nadir Values and Their Relative Errors under Different $N$}
                \vspace{-2ex}
                \label{tab:accuracy}
                \centering
                \newcommand{\tabincell}[2]{\begin{tabular}{@{}#1@{}}#2\end{tabular}}
                \begin{tabular}{cccccc}
                    \toprule
                     & \multirow{2}{*}{Simulink} & \multicolumn{4}{c}{BP approximation}\\
                    \cmidrule{3-6}
                     &  & $N=1$ & $N=2$ & $N=4$ & $N=8$\\
                    \midrule
                    Nadir value (Hz) & \hspace{-1ex}$-0.3884$\hspace{-1ex} & \hspace{-1ex}$-0.3759$\hspace{-1ex} & \hspace{-1ex}$-0.3893$\hspace{-1ex} & \hspace{-1ex}$-0.3892$\hspace{-1ex} & \hspace{-1ex}$-0.3883$\hspace{-1ex}\\
                    Relative error & N/A & $3.21\%$ & $0.23\%$ & $0.20\%$ & $0.02\%$\\
                    \bottomrule
                \end{tabular}
                \vspace{-3ex}
            \end{table}

            We consider a power imbalance of 20MW (about 10\% of the total load) and the frequency dynamics in the subsequent horizon $H$.
            Fig.~\ref{fig:accuracy-even}(a) depicts the frequency dynamics obtained from the BP approximation and those obtained from numerical simulation (by Simulink).
            In this experiment, we vary the length of $H$ from 5s to 30s and keep the segment number $N=1$.
            We report the estimated nadir values and their relative errors in Table~\ref{tab:accuracy-horizon}.
            From Fig.~\ref{fig:accuracy-even}(a) and Table~\ref{tab:accuracy-horizon}, we observe that when $H=5$s, the BP approximation performs well in fitting frequency dynamics.
            However, in this case, $H=5$s is too short for frequency dynamics to reach the nadir, which results in the 2.73\% error.
            Instead, $H=10$s is long enough to reach the nadir, and hence, the relative error reduces to 0.20\%.
            However, before an event occurs, system operators do not know when the nadir will take place and thus may select a relatively longer $H$ to cover the nadir.
            In this case, as $H$ gets longer, the relative error of nadir increases quickly.
            This is because the 1-segment BP approximation is not able to fit frequency dynamics for infinitely long.
            To tackle this issue, we divide $H$ evenly into $N$ segments and compare the results with different $N$.
            We use the 30s-horizon for analysis and report the obtained frequency dynamics and the estimated nadir values in Fig.~\ref{fig:accuracy-even}(b) and Table~\ref{tab:accuracy}, respectively.
            Our results show that, as $N$ increases, each segment shortens and the performance of BP quickly improves.

            We provide a discussion on how to set up the BP approximation in a general condition.
            According to the above results, we notice that dividing the horizon $H$ into multiple segments and applying a multi-segment BP approximation produce superb performance.
            On the other hand, in view of \eqref{eq:dynamic-SST}--\eqref{eq:reserve-SST}, the number of constraints in the corresponding UC formulation increases linearly with the number of segments.
            Consequently, we need to strike a right balance between approximation performance and computational burden.
            In addition, because the frequency nadir is the critical point of the dynamics, the BP segments before the nadir and covering the nadir is more important than those afterwards.
            Hence, we suggest an uneven division of $H$, in which the early segments should be shorter for a more accurate estimate of $\Delta f(t)|_{\text{nadir}}$ and the latter segments should be longer to relieve computational burden.

\begin{figure}[!htbp]	
                \centering
                \subfigure[6-bus system]{\includegraphics[width=0.45\textwidth]{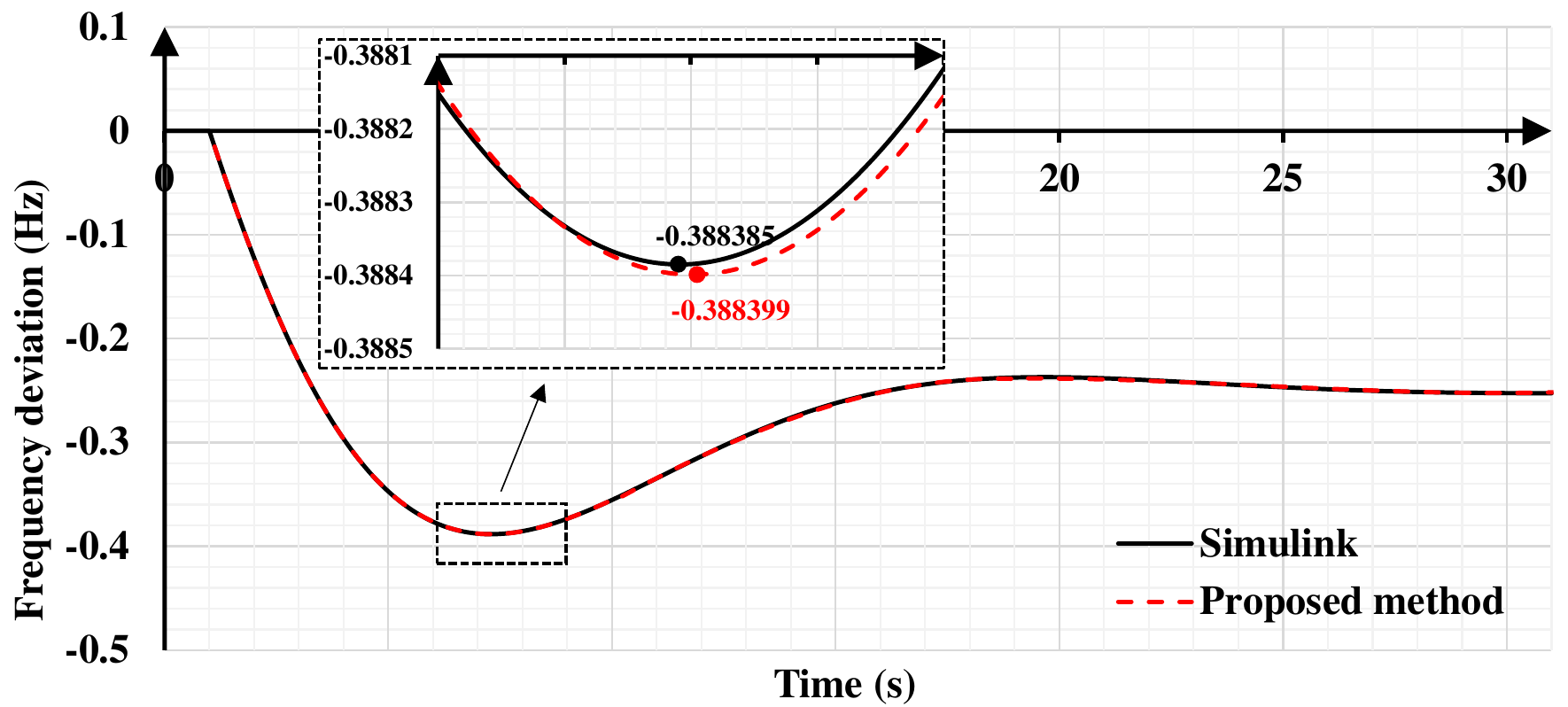}}\\
                \vspace{-1ex}
                \subfigure[118-bus system]{\includegraphics[width=0.45\textwidth]{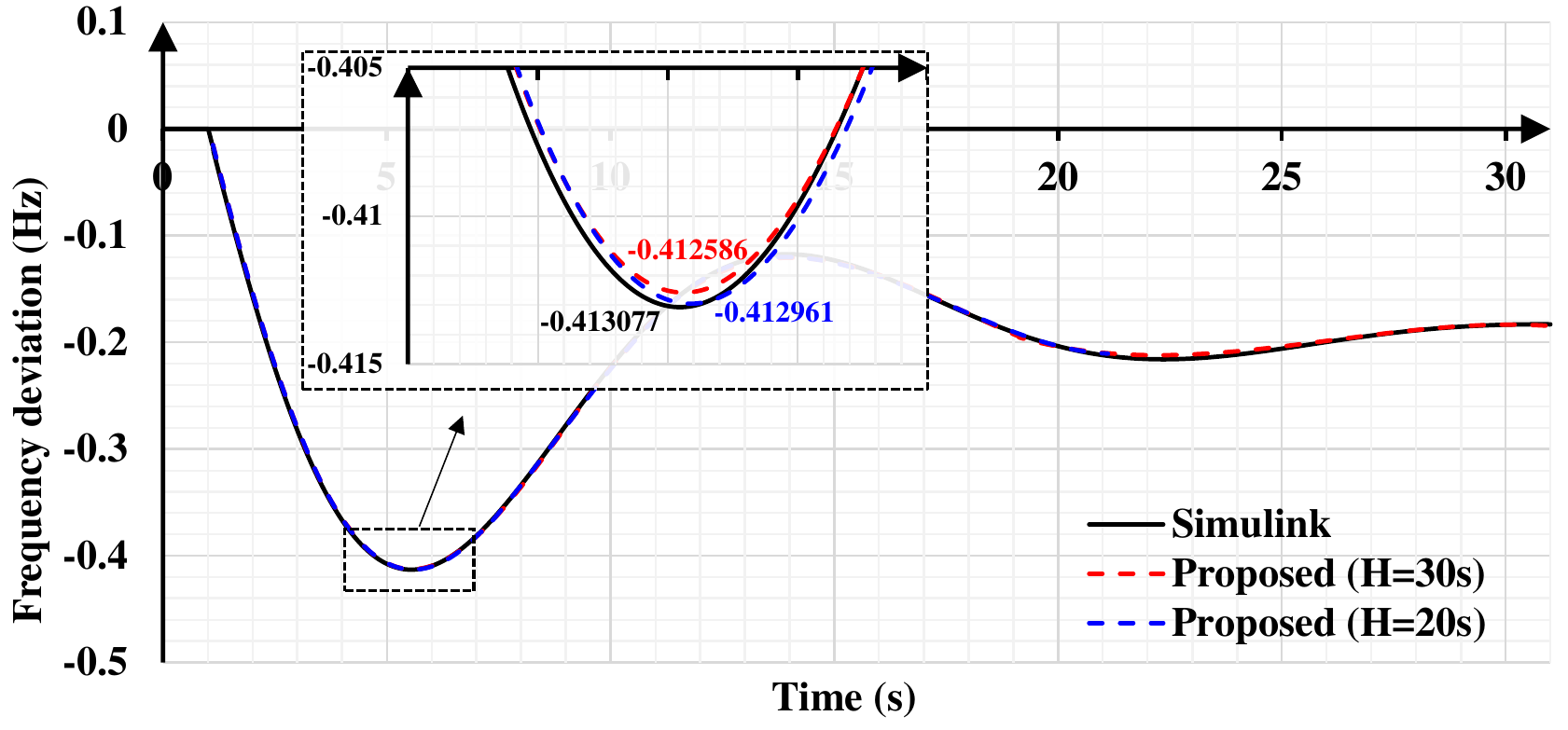}}
                \caption{BP approximation with an uneven division of the horizon}	
                \label{fig:accuracy-uneven}
            \end{figure}

            Along the above discussion, we divide the 30s-horizon into 4 segments with 10\%, 20\%, 30\%, and 40\% of the horizon, respectively.
            We depict the frequency dynamics approximation in this case in Fig.~\ref{fig:accuracy-uneven}(a).
            From this figure, we observe that the BP approximation is highly accurate and the relative error in estimating $\Delta f(t)|_{\text{nadir}}$ is lower than 0.004\%.
            In addition, we conduct an accuracy validation in the IEEE 118-bus system, and the system parameters will be given in Section \ref{sec:results}.C.
            We show the results in Fig.~\ref{fig:accuracy-uneven}(b).
            The BP approximation remains highly accurate in the 118-bus system and the relative error is lower than 0.12\%.
            Moreover, if we set $H=20$s, the relative error can be further reduced to lower than 0.03\%.
            These results validate the tightness of the proposed method.

\begin{figure}[!htbp]	
                \centering
                \vspace{-3ex}
                \includegraphics[width=0.45\textwidth]{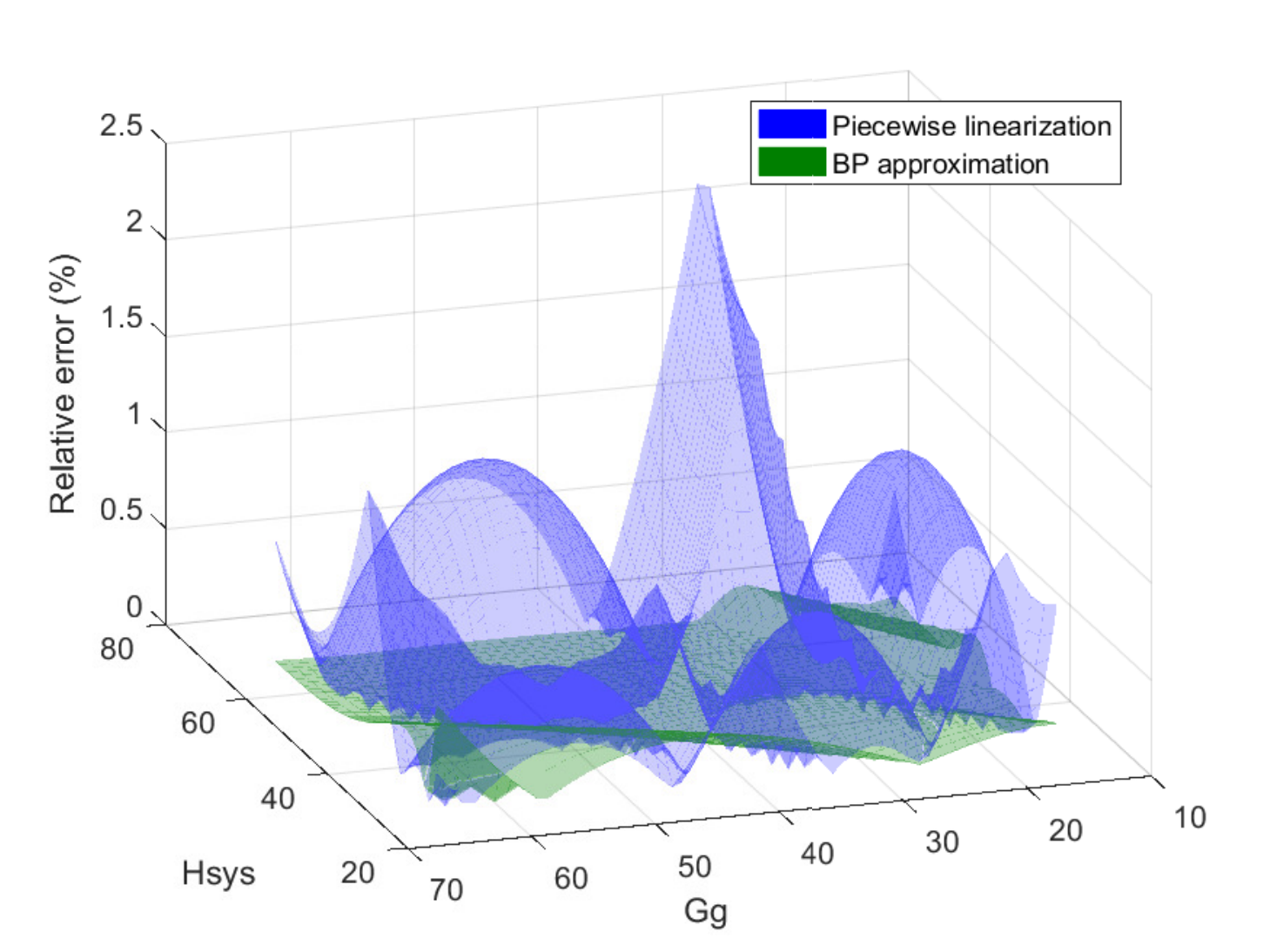}
                \vspace{-2ex}
                \caption{Relative error in estimating $\Delta f_{\text{nadir}}$}	
                \label{fig:accuracy comparison}
            \end{figure}

            We also compare our BP approximation with the piecewise linearization used in \cite{ZhangZhou-3475, LiQiao-1435, AhmadiGhasemi-4685, ZhangDu-4684,   LiuHu-5670, ZhangWu-5669}.
            Because the piecewise linearization method ignores the dead band and requires the response constant of all units to be the same, it is inapplicable to the case shown in Fig.  \ref{fig:SFR}.
            For comparison, we do not consider the dead band and modify all response constants to be 7s.
            We vary the total inertia $H_{sys}$ and the sum of the droop factors of units $G_{g}$ and obtain 99 evaluation points.
            4 segments are considered for piecewise linearization.
            Fig. \ref{fig:accuracy comparison} compares the relative error of piecewise linearization and that of BP approximation in estimating nadir frequency deviation.
            TABLE \ref{tab:accuracy comparison} provides the average relative error of the two methods.
            We can see that BP approximation performs better than piecewise linearization in most cases and the average relative error is one order of magnitude lower than that of piecewise linearization.
            The results validate the higher accuracy of BP approximation compared to existing methods.
            \begin{table}[!htbp]
                \renewcommand{\arraystretch}{1.1}
                \vspace{-1ex}
                \caption{Average Relative Errors in Estimating $\Delta f_{\text{nadir}}$}
                \vspace{-2ex}
                \label{tab:accuracy comparison}
                \centering
                \newcommand{\tabincell}[2]{\begin{tabular}{@{}#1@{}}#2\end{tabular}}
                \begin{tabular}{ccc}
                    \toprule
                     & Piecewise Linearization & BP approximation\\
                    \midrule
                    Average relative error & 0.5031\% & 0.0653\%\\
                    Highest relative error & 2.2734\% & 0.5543\%\\
                    \bottomrule
                \end{tabular}
                \vspace{-3ex}
            \end{table}

        \vspace{-0ex}
        \subsection{6-Bus System} \label{sec:6-bus}
\begin{figure}[!t]	
                \centering
                \includegraphics[width=0.45\textwidth]{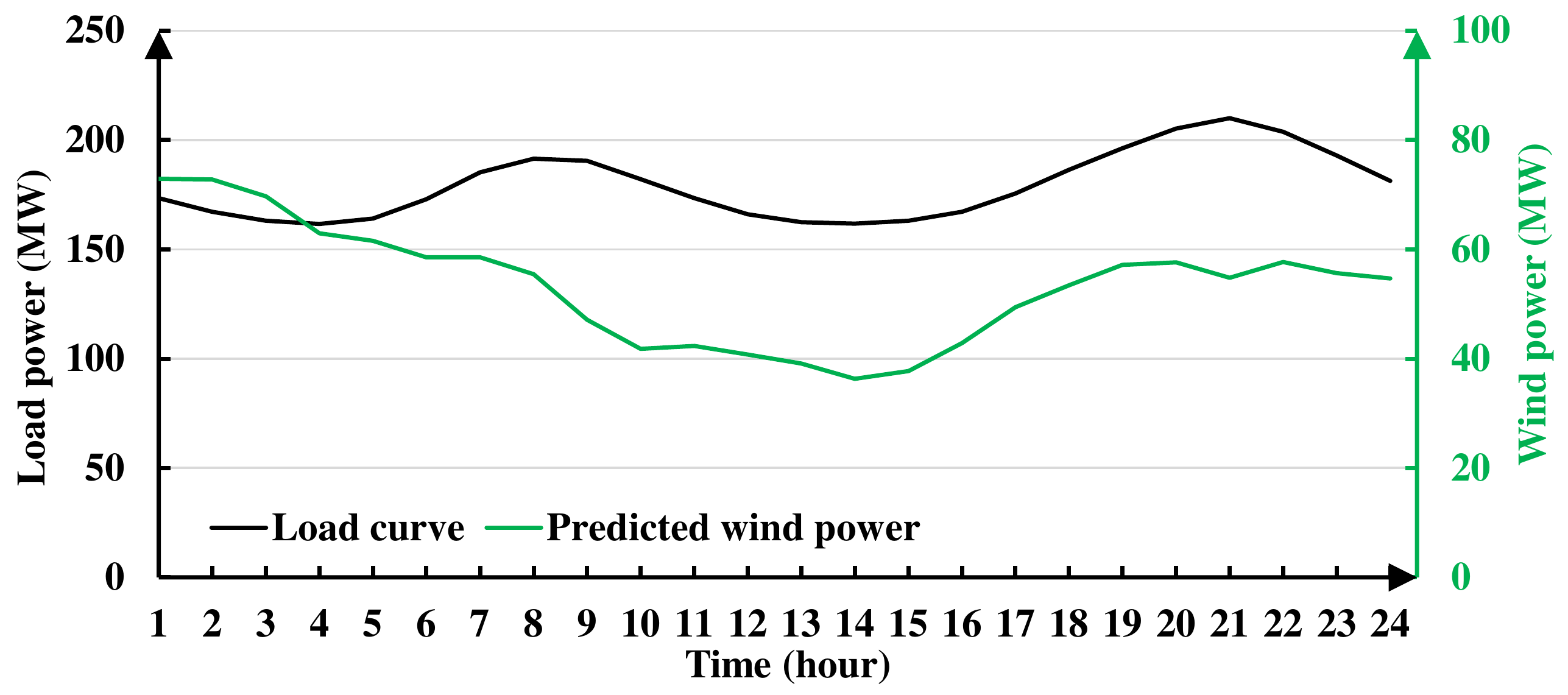}
                \vspace{-1ex}
                \caption{Load curve and predicted wind power}	
                \vspace{-1ex}
                \label{fig:curves}
            \end{figure}

            Next, we use the IEEE 6-bus system for analysis.
            The system includes 3 thermal units (G1, G2, and G3) and a newly-added wind farm (W).
            The load curve and the predicted available wind power are depicted in Fig.~\ref{fig:curves}.
            The maximum total load is 210MW.
            The standard deviation of wind power prediction is set as 5\% of the predicted value.
            The risk level $\epsilon$ in chance constraints is set as 10\% and the radius $r$ of the Wasserstein ball for DRCC is set as 0.01.

            Table~\ref{tab:parameter} displays the main technical parameters of thermal units and the wind farm.
            The fuel cost coefficients are the same as in~\cite{ZhouFang-3859}.
            The droop factor of the wind farm is set as 20 MW/Hz by default and can be adjusted from 10 MW/Hz to 25 MW/Hz with a stepsize of 5 MW/Hz.
            The nominal frequency is 50 Hz and the dead band is set as 0.015 Hz~\cite{YangXu-3622}.
            The cost coefficients of PFR reserves are 15 \$/MWh and 5 \$/MWh for thermal units and the wind farm, respectively.
            In each hour, we consider a sudden power imbalance of 10\% total load \cite{WenLi-4681}.
            We set the load damping factor as 1\% total load per Hz, the maximum admitted RoCoF as 0.5 Hz/s, and the maximum admitted frequency deviation at the nadir and at the QSS as 0.5 Hz and 0.3 Hz, respectively \cite{LiQiao-1435}.
            We consider a 30s-horizon of frequency dynamics and divide it into 4 segments with 10\%, 20\%, 30\%, and 40\% of the horizon, respectively.
            \begin{table}[!t]
                \renewcommand{\arraystretch}{1.0}
                \vspace{-0ex}
                \caption{Main Technical Parameters of the 6-Bus System}
                \vspace{-2ex}
                \label{tab:parameter}
                \centering
                \newcommand{\tabincell}[2]{\begin{tabular}{@{}#1@{}}#2\end{tabular}}
                \begin{tabular}{ccccc}
                    \toprule
                    Units & G1 & G2 & G3 & W\\
                    \midrule
                    Capacity (MW) & 200 & 150 & 180 & 80\\
                    Minimum generation (MW) & 60 & 45 & 54 & N/A\\
                    Inertia constant (s) & 8 & 5 & 6 & 5\\
                    Response constant (s) & 10 & 4 & 6 & 0\\
                    Droop factor (MW/Hz) & 20 & 25 & 18 & 10$\sim$25\\
                    \bottomrule
                \end{tabular}
                \vspace{-1ex}
            \end{table}

        \subsubsection{Baseline Results}
            With the above parameter settings, we compare the following two UC formulations:

            (\textbf{Case 1}) without frequency security constraints; and

            (\textbf{Case 2}) with frequency security constraints~\eqref{eq:security} and~\eqref{eq:frequency}.

\begin{figure}[!htbp]	
                \centering
                \vspace{-0ex}
                \includegraphics[width=0.45\textwidth]{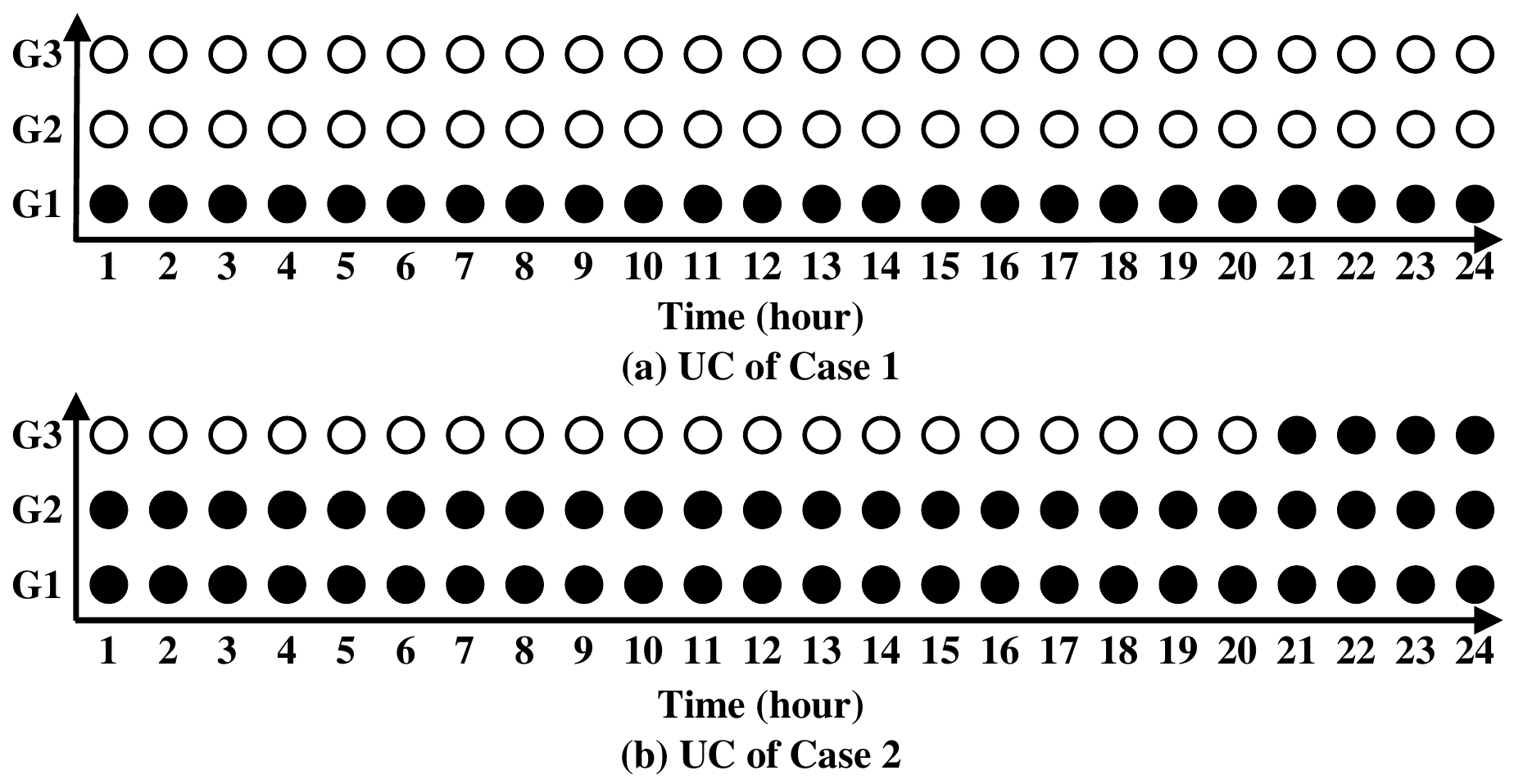}
                \vspace{-2ex}
                \caption{Optimal UC in Cases 1 and 2}	
                \vspace{-2ex}
                \label{fig:UC12}
            \end{figure}

\begin{figure}[!htbp]	
                \centering
                \subfigure[Initial RoCoF]{\includegraphics[width=0.43\textwidth]{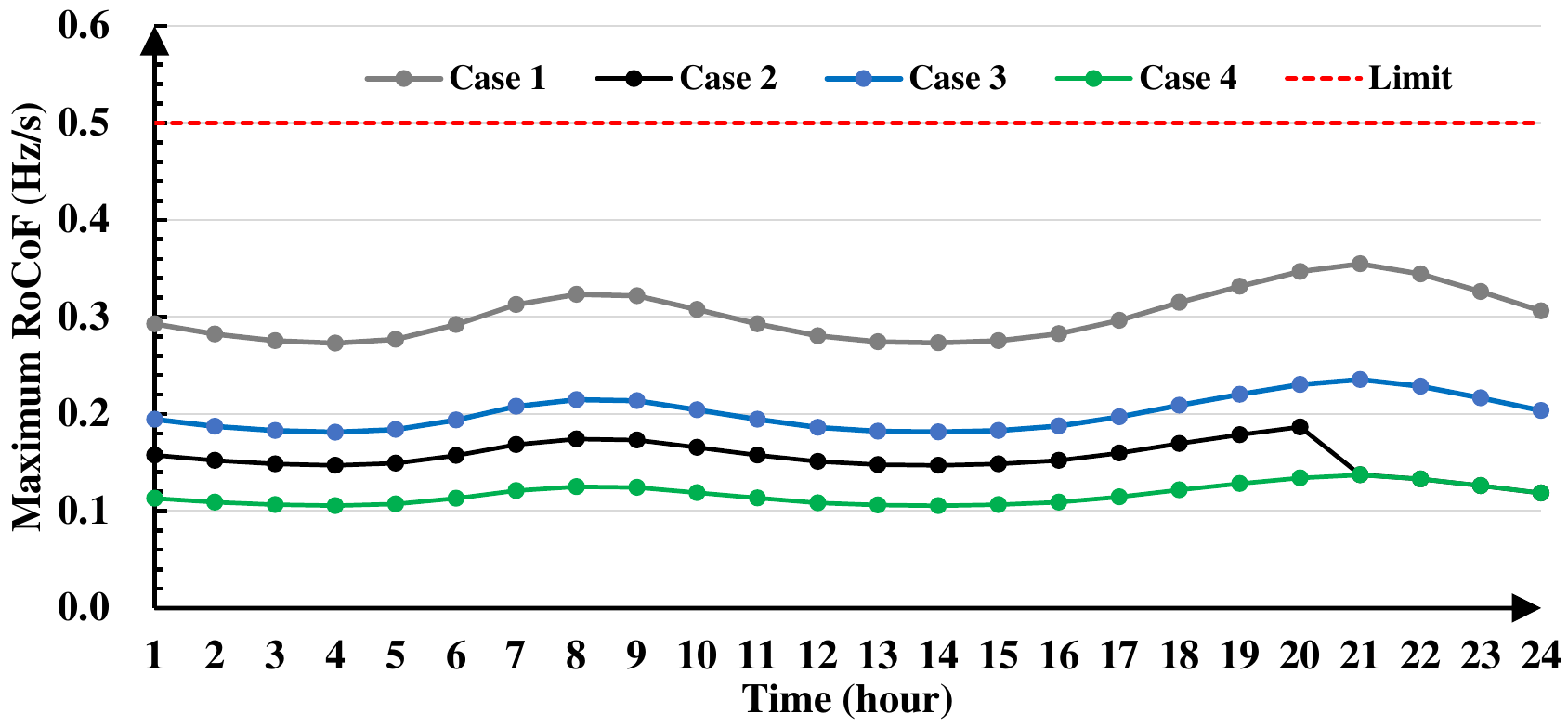}}\\
                \subfigure[Nadir deviation]{\includegraphics[width=0.43\textwidth]{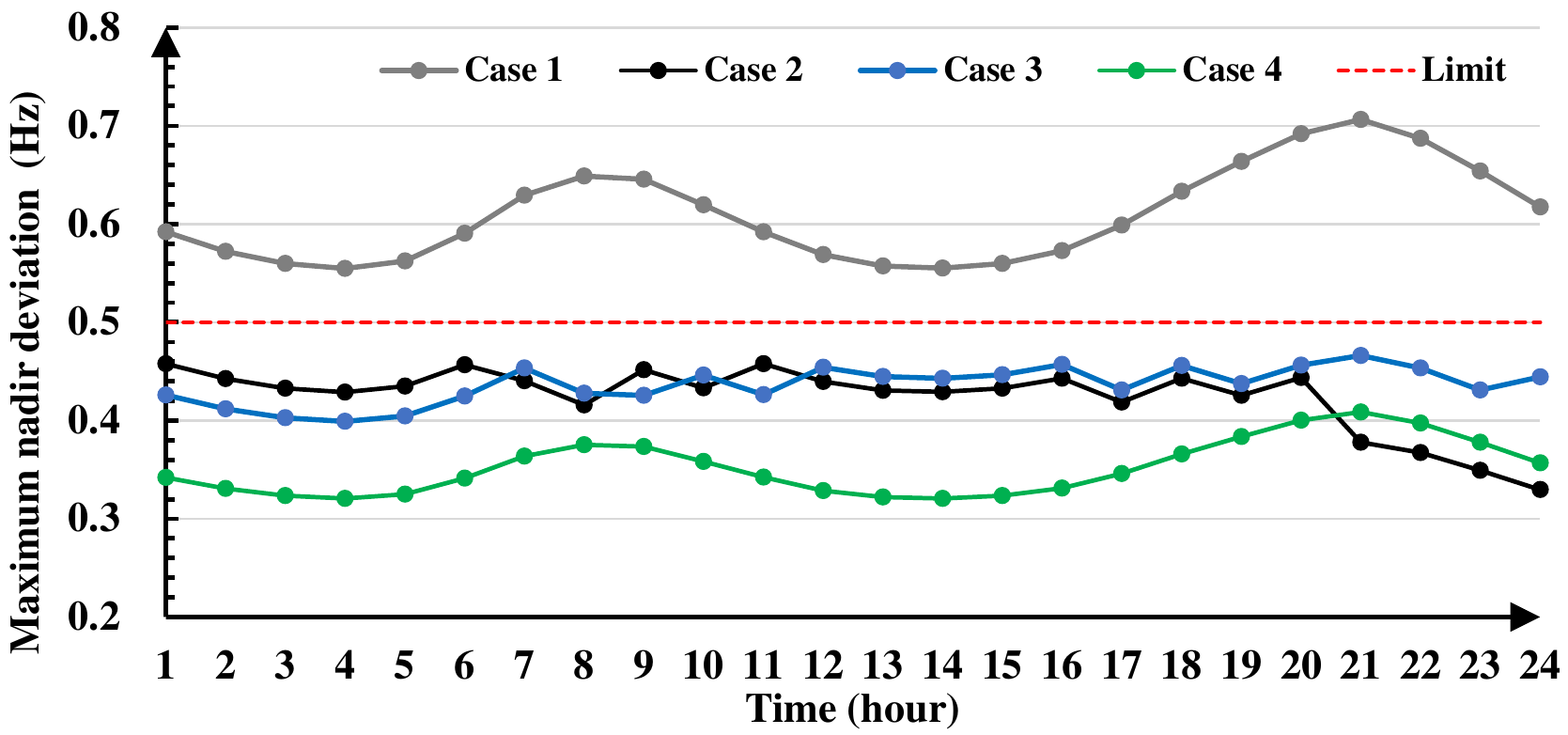}}
                \vspace{-1ex}
                \caption{Initial RoCoF and nadir deviation at different periods}	
                \label{fig:transient}
            \end{figure}

            Note that in Case 1, we do not consider frequency security constraints but require the total PFR reserve to be no less than the set power imbalance.
            We report the optimal UC solutions and frequency security metrics of these two cases in Figs.~\ref{fig:UC12} and \ref{fig:transient}, respectively, as well as their performance details in Table~\ref{tab:results12}.

\begin{figure}[!htbp]	
                \centering
                \vspace{-2ex}
                \includegraphics[width=0.43\textwidth]{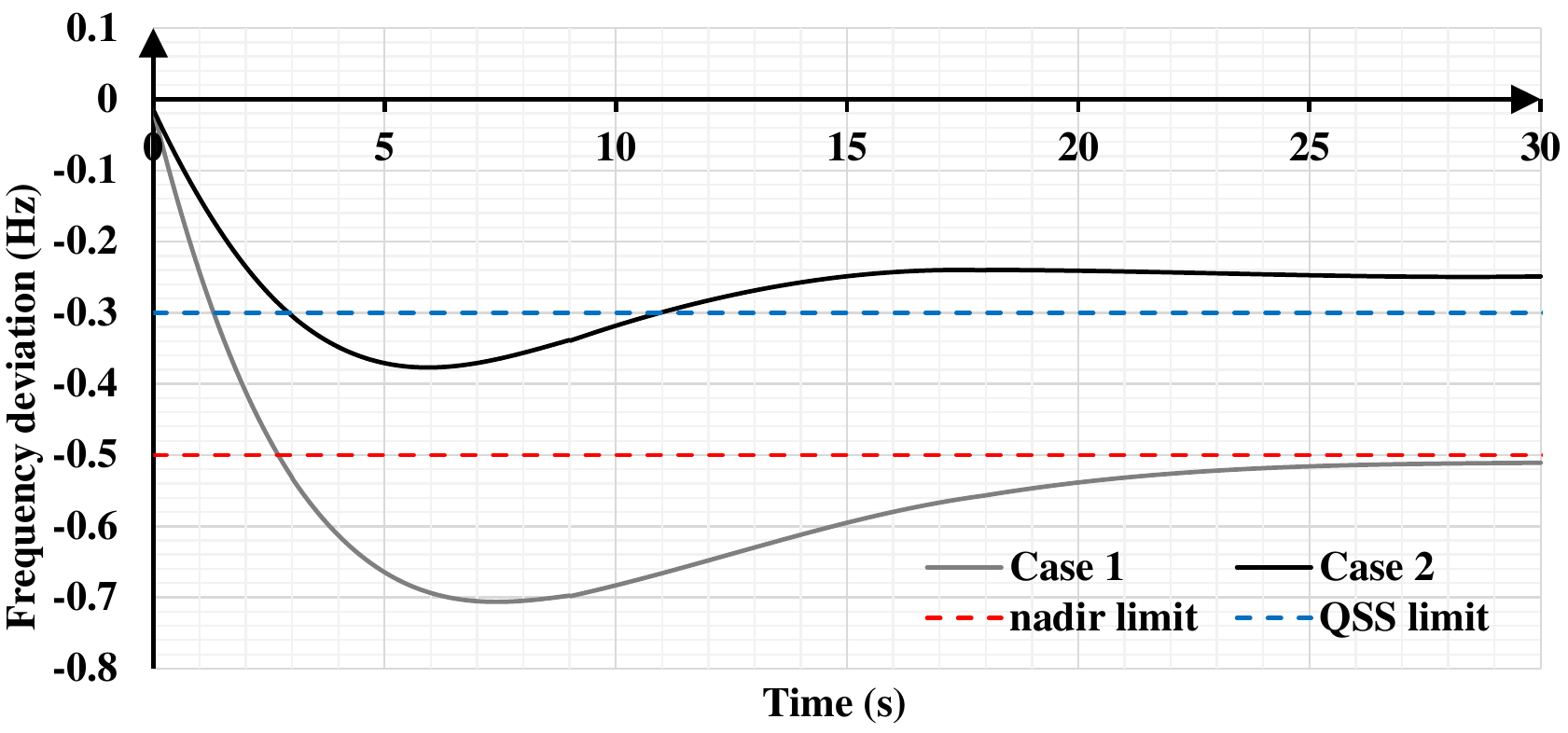}
                \vspace{-1ex}
                \caption{Frequency dynamics of hour 21 in Cases 1 and 2}	
                \vspace{-1ex}
                \label{fig:dynamics-case12}
            \end{figure}

            From Fig.~\ref{fig:UC12} and Table~\ref{tab:results12}, we observe that in Case 1, only G1 is online.
            For this case, we simulate the frequency dynamics under power imbalance and find that in hour 21, the frequency deviation reaches 0.7133 Hz at the nadir and reaches 0.5131 Hz at the QSS, which exceeds the frequency security limits (by 0.5 Hz and 0.3 Hz, respectively). The frequency dynamics of hour 21 in Case 1 are provided in Fig.~\ref{fig:dynamics-case12}.

            \begin{table}[!htbp]
                \renewcommand{\arraystretch}{1.0}
                \vspace{-0ex}
                \caption{Detailed Comparison Between Cases 1 and 2}
                \vspace{-2ex}
                \label{tab:results12}
                \centering
                \newcommand{\tabincell}[2]{\begin{tabular}{@{}#1@{}}#2\end{tabular}}
                \begin{tabular}{ccc}
                    \toprule
                    Comparison terms & Case 1 & Case 2\\
                    \midrule
                    Start-up \& shut-down cost (\$) & 0 & 5400\\
                    Fuel cost of thermal units (\$) & 43763.09 & 49040.87\\
                    PFR reserve cost of thermal units (\$) & 6445.92 & 5171.99\\
                    PFR reserve cost of wind farm (\$) & 0 & 919.69\\
                    Total cost (\$) & 50209.01 & 60532.55\\
                    \midrule
                    Total PFR reserve of thermal units (MWh) & 429.73 & 344.80\\
                    Total PFR reserve of wind farm (MWh) & 0 & 183.94\\
                    \midrule
                    Maximum RoCoF (Hz/s) & 0.2625 & 0.1866\\
                    Maximum $\Delta f$ at the nadir (Hz) & 0.7133 & 0.4578\\
                    Maximum $\Delta f$ at the quasi-steady state (Hz) & 0.5131 & 0.2995\\
                    \midrule
                    Computation time (s) & 8.61 & 38.42\\
                    \bottomrule
                \end{tabular}
                \vspace{-3ex}
            \end{table}


\begin{figure}[!t]	
                \centering
                \vspace{-0ex}
                \includegraphics[width=0.45\textwidth]{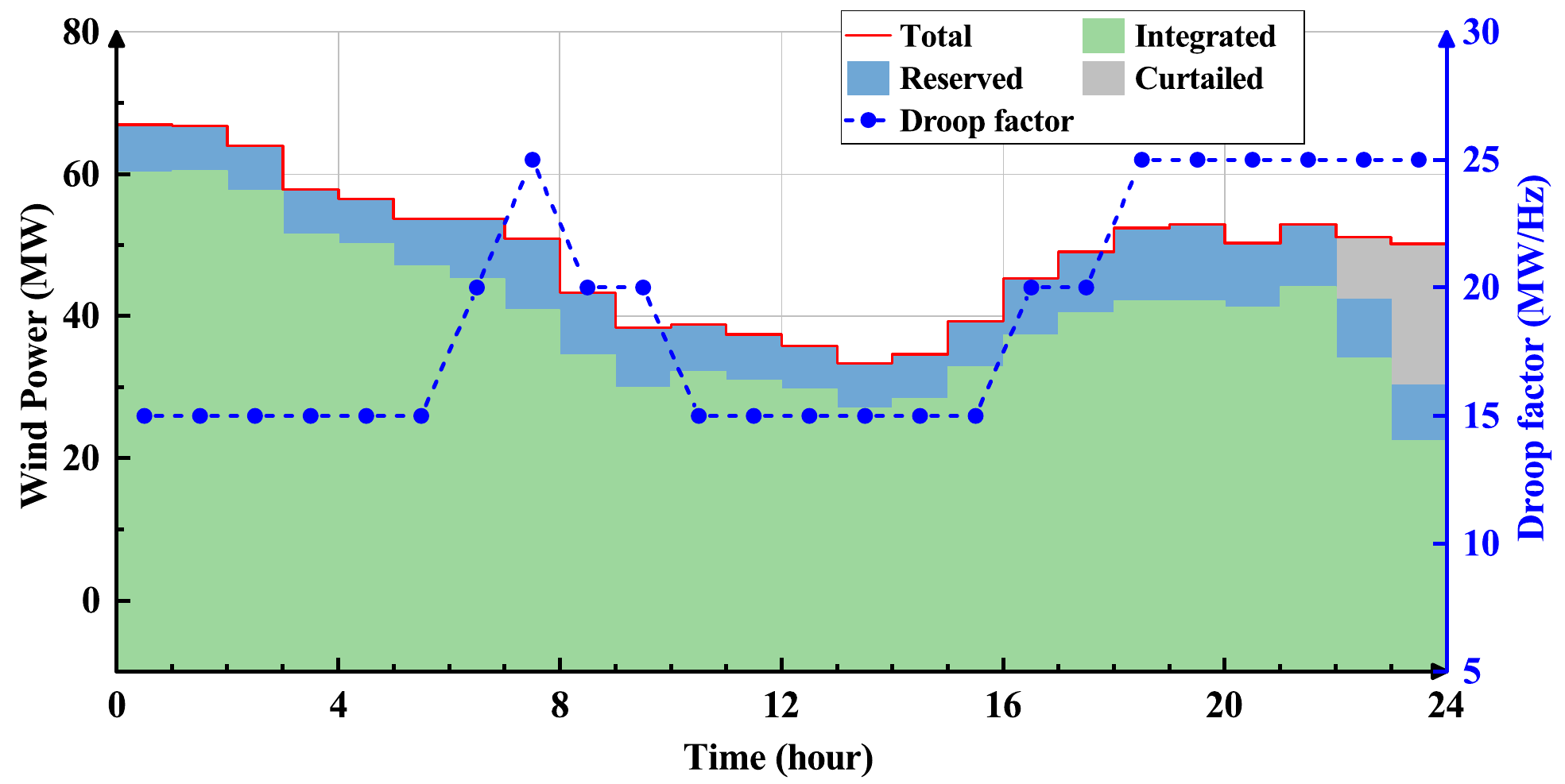}
                \vspace{-1ex}
                \caption{Operation of the wind farm in Case 2}	
                \vspace{-0ex}
                \label{fig:wind-case2}
            \end{figure}

\begin{figure}[!t]	
                \centering
                \includegraphics[width=0.45\textwidth]{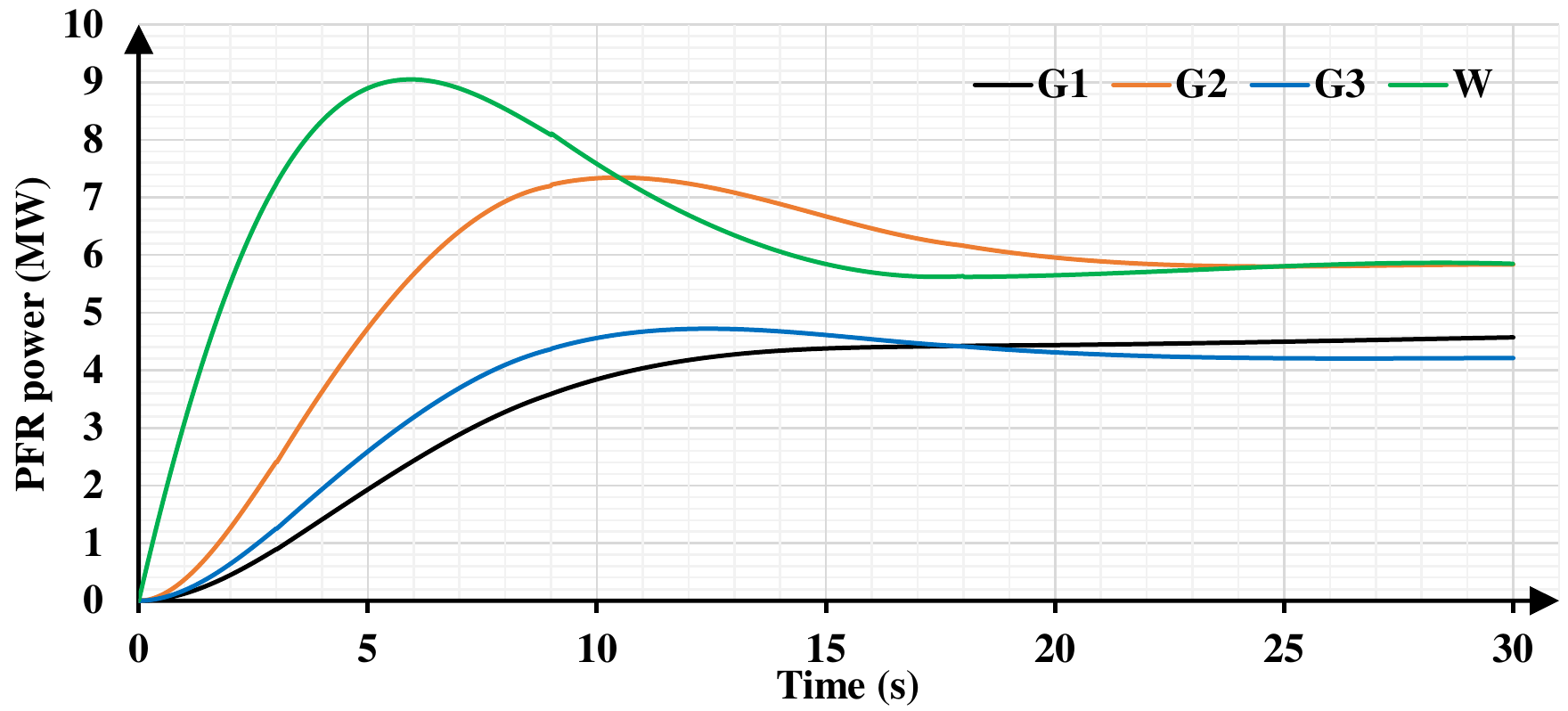}
                \vspace{-1ex}
                \caption{PFR power dynamics of hour 21 in Case 2}	
                \vspace{-2ex}
                \label{fig:PFR-case2}
            \end{figure}

            In contrast, in Case 2, G1 and G2 are always online and G3 is started up in hour 21, which provides more inertia and PFR capacity than those in Case 1.
            We depict the PFR reserve and the droop factor of the wind farm in Fig.~\ref{fig:wind-case2}.
            From this figure, we observe that most wind power is integrated for power supply or reserved for frequency support, and a only small portion of wind power is curtailed in the last two hours. In addition, the droop factor varies across different hours, which will be further discussed later.
            As a consequence of more units online and more PFR reserves from both the thermal units and the wind farm, in Case 2, the maximum frequency deviation becomes 0.4578 Hz at the nadir and 0.2995 Hz at the QSS, significantly lower than those in Case 1.
            Note that, since G3 is online from hour 21 in Case 2, the maximum nadir frequency deviation occurs in hour 20 and the maximum QSS frequency deviation occurs in hour 11.
            We also depict the frequency dynamics of hour 21 in Case 2 in Fig.~\ref{fig:dynamics-case12} for comparison and the corresponding PFR power dynamics in Fig.~\ref{fig:PFR-case2}.
            Because the unit commitment is discrete and the status change of any unit will strongly affect frequency dynamics, it is hard to find a unit commitment that is just right to satisfy all frequency security limits, for which the frequency dynamics of Case 2 in Fig.~\ref{fig:dynamics-case12} seem conservative.
            The above analyses demonstrate that the proposed model can effectively protect frequency security.

        \subsubsection{Impact of the Dead Band}
            To evaluate the value of modeling the dead band in frequency security, we consider a third UC formulation:\\
            (\textbf{Case 3}) without considering the dead band in Case 2.

\begin{figure}[!t]	
                \centering
                \vspace{-0ex}
                \includegraphics[width=0.45\textwidth]{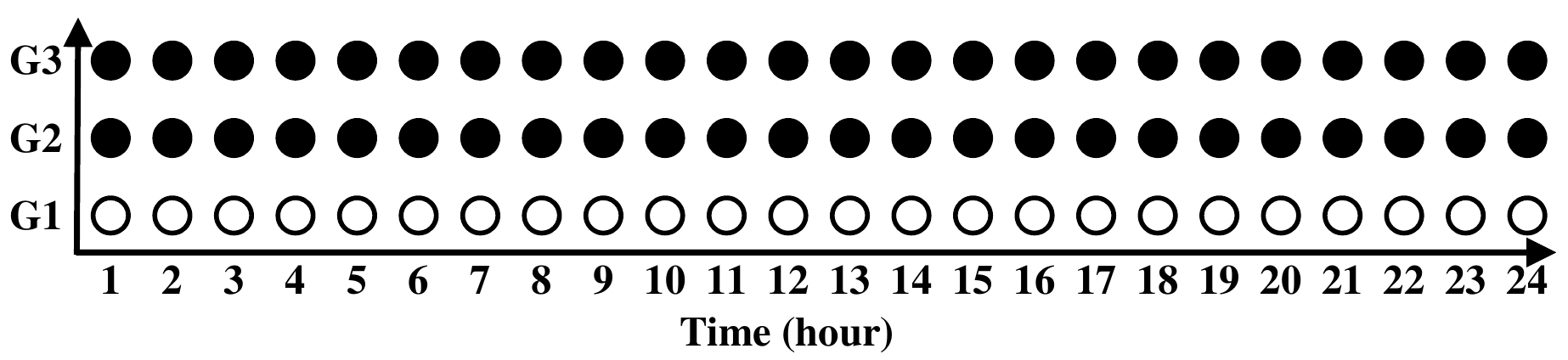}
                \vspace{-1ex}
                \caption{Optimal UC in Case 3}	
                \vspace{-2ex}
                \label{fig:UC3}
            \end{figure}

            We report the optimal UC solution in Case 3 in Fig.~\ref{fig:UC3} and the performance details in Table~\ref{tab:results3}.
            From Fig.~\ref{fig:UC3} and Table~\ref{tab:results3}, we observe that in Case 3, G2 and G3 remain online while G1 remains offline across all periods.
            Because the more economic G2 and G3 are utilized, the total cost in Case 3 is lower than that in Case 2.
            However, ignoring the dead band causes undesirable frequency dynamics.
            From Table~\ref{tab:results3}, we observe that the maximum $\Delta f(t)|_{\text{QSS}}$ in Case 3 reaches 0.3141 Hz, exceeding frequency security limit (0.3 Hz).
            This highlights the necessity of considering the dead band in frequency stability-constrained UC.

            \begin{table}[!t]
                \renewcommand{\arraystretch}{1.0}
                \vspace{-0ex}
                \caption{Detailed Comparison Between Cases 3 and 2}
                \vspace{-2ex}
                \label{tab:results3}
                \centering
                \newcommand{\tabincell}[2]{\begin{tabular}{@{}#1@{}}#2\end{tabular}}
                \begin{tabular}{ccc}
                    \toprule
                    Comparison terms & Case 3 & Case 2\\
                    \midrule
                    Start-up \& shut-down cost (\$) & 0 & 5400\\
                    Fuel cost of thermal units (\$) & 47718.70 & 49040.87\\
                    PFR reserve cost of thermal units (\$) & 4712.09 & 5171.99\\
                    PFR reserve cost of wind farm (\$) & 1033.06 & 919.69\\
                    Total cost (\$) & 53463.85 & 60532.55\\
                    \midrule
                    Maximum RoCoF (Hz/s) & 0.2354 & 0.1866\\
                    Maximum $\Delta f$ at the nadir (Hz) & 0.4662 & 0.4578\\
                    Maximum $\Delta f$ at the quasi-steady state (Hz) & 0.3141 & 0.2995\\
                    \midrule
                    Computation time (s) & 57.15 & 38.42\\
                    \bottomrule
                \end{tabular}
                \vspace{-0ex}
            \end{table}

        \subsubsection{Benefits of Variable Droop Factors}
            To evaluate the benefit of adopting variable droop factors of the wind farm, as opposed of constant droop factors, we consider a fourth UC formulation:

            (\textbf{Case 4}) fixing the droop factor in Case 2 at 20 MW/Hz.

\begin{figure}[!t]	
                \centering
                \vspace{1ex}
                \includegraphics[width=0.45\textwidth]{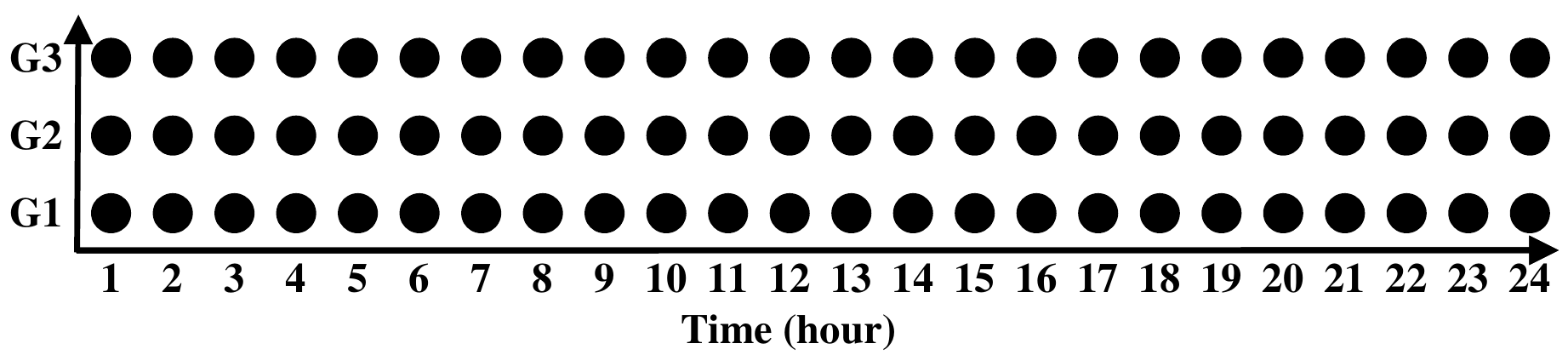}
                \vspace{-1ex}
                \caption{Optimal UC in Case 4}	
                \vspace{-2ex}
                \label{fig:UC4}
            \end{figure}

            We report the optimal UC solution in Case 4 in Fig.~\ref{fig:UC4} and the performance details in Table~\ref{tab:results4}.
            \begin{table}[!t]
                \renewcommand{\arraystretch}{1.0}
                \vspace{-0ex}
                \caption{Detailed Comparison Between Cases 4 and 2}
                \vspace{-2ex}
                \label{tab:results4}
                \centering
                \newcommand{\tabincell}[2]{\begin{tabular}{@{}#1@{}}#2\end{tabular}}
                \begin{tabular}{ccc}
                    \toprule
                    Comparison terms & Case 4 & Case 2\\
                    \midrule
                    Start-up \& shut-down cost (\$) & 0 & 5400\\
                    Fuel cost of thermal units (\$) & 59198.07 & 49040.87\\
                    PFR reserve cost of thermal units (\$) & 5595.82 & 5171.99\\
                    PFR reserve cost of wind farm (\$) & 809.83 & 919.69\\
                    Total cost (\$) & 65603.72 & 60532.55\\
                    \midrule
                    Maximum RoCoF (Hz/s) & 0.1371 & 0.1866\\
                    Maximum $\Delta f$ at the nadir (Hz) & 0.4087 & 0.4578\\
                    Maximum $\Delta f$ at the quasi-steady state (Hz) & 0.2614 & 0.2995\\
                    \midrule
                    Computation time (s) & 4.89 & 38.42\\
                    \bottomrule
                \end{tabular}
                \vspace{-0ex}
            \end{table}
            \begin{figure}[!t]	
                \centering
                \vspace{-0ex}
                \includegraphics[width=0.45\textwidth]{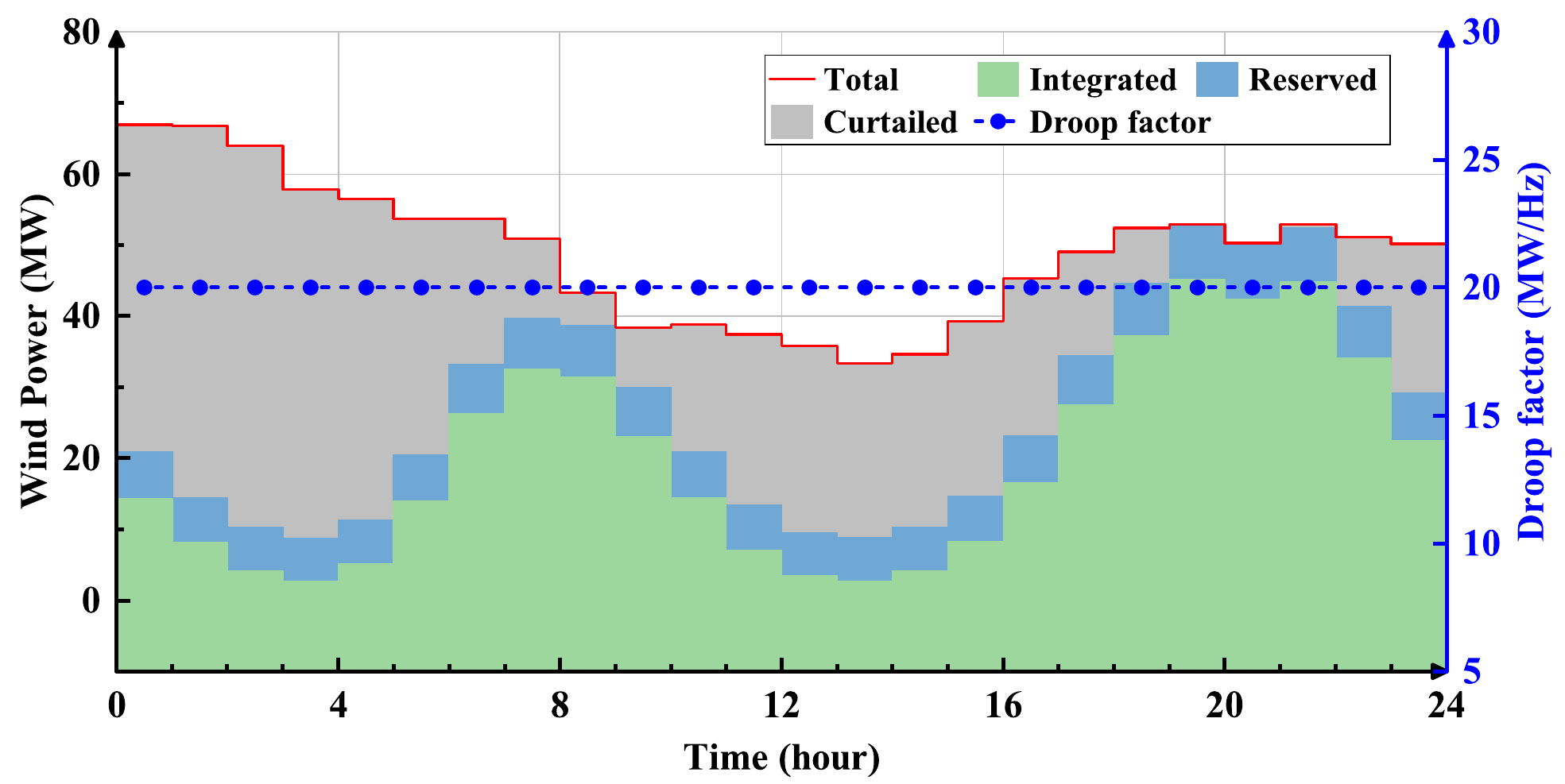}
                \vspace{-2ex}
                \caption{Operation of the wind farm in Case 4}	
                \vspace{-0ex}
                \label{fig:wind-case4}
            \end{figure}

            From Fig.~\ref{fig:UC4} and Table~\ref{tab:results4}, we observe that in Case 4, all thermal units remain online all the time and the total cost is higher than that in Case 2.
            We provide the PFR reserve and droop factors of the wind farm in Case 4 in Fig.~\ref{fig:wind-case4}.
            Comparing Figs.~\ref{fig:wind-case4} and~\ref{fig:wind-case2}, we observe that much more wind curtailment occurs in Case 4.
            The reason is analyzed in the following.
            In Case 4, if we replace its UC with that in Case 2 while keeping the droop factors unchanged, $\Delta f(t)|_{\text{QSS}}$ would reach 0.3007 Hz at hour 8 and 0.3207 Hz at hour 20, which exceed the frequency security limit (0.3 Hz).
            Hence, in Case 4, G3 needs to be online to reduce $\Delta f(t)|_{\text{QSS}}$, and consequently, much wind power needs to be curtailed to satisfy the minimum generation limits of the thermal units.
            In contrast, in Case 2, we reduce $\Delta f(t)|_{\text{QSS}}$ by increasing the droop factor (such as in hours 8 and 20; see Fig.~\ref{fig:wind-case2}), which avoids putting G3 online and reduces the total cost by 7.69\%.
            These results demonstrate that variable droop factors can improve the coordination of different frequency resources for higher operation economy.

        \subsubsection{Comparison with Existing Methods}
            To compare our method with the piecewise linearization method in existing works \cite{ZhangZhou-3475, LiQiao-1435, AhmadiGhasemi-4685, ZhangDu-4684,   LiuHu-5670, ZhangWu-5669}, we consider a fifth UC formulation:

            (\textbf{Case 5}) replacing BP approximation in Case 3 by piecewise linearization.

            There are some notes for the comparison.
            First, due to the assumptions of the piecewise linearization method, we do not consider the dead band and modify all response constants to be 7s.
            Accordingly, we compare Case 5 with Case 3.
            Second, since it is hard to derive an accurate piecewise linearization, to relieve the computational burden, we only use the possible combinations of UC and $G_{w}$ as evaluate points.
            Deriving the linearization consumes about 600s.
            Third, the piecewise linearization method does not calculate the maximum PFR power during frequency dynamics, so we choose a conservative PFR reserve according to droop factors and $\overline{\Delta f_{err}}$.

\begin{figure}[!t]	
                \centering
                \vspace{-1ex}
                \includegraphics[width=0.45\textwidth]{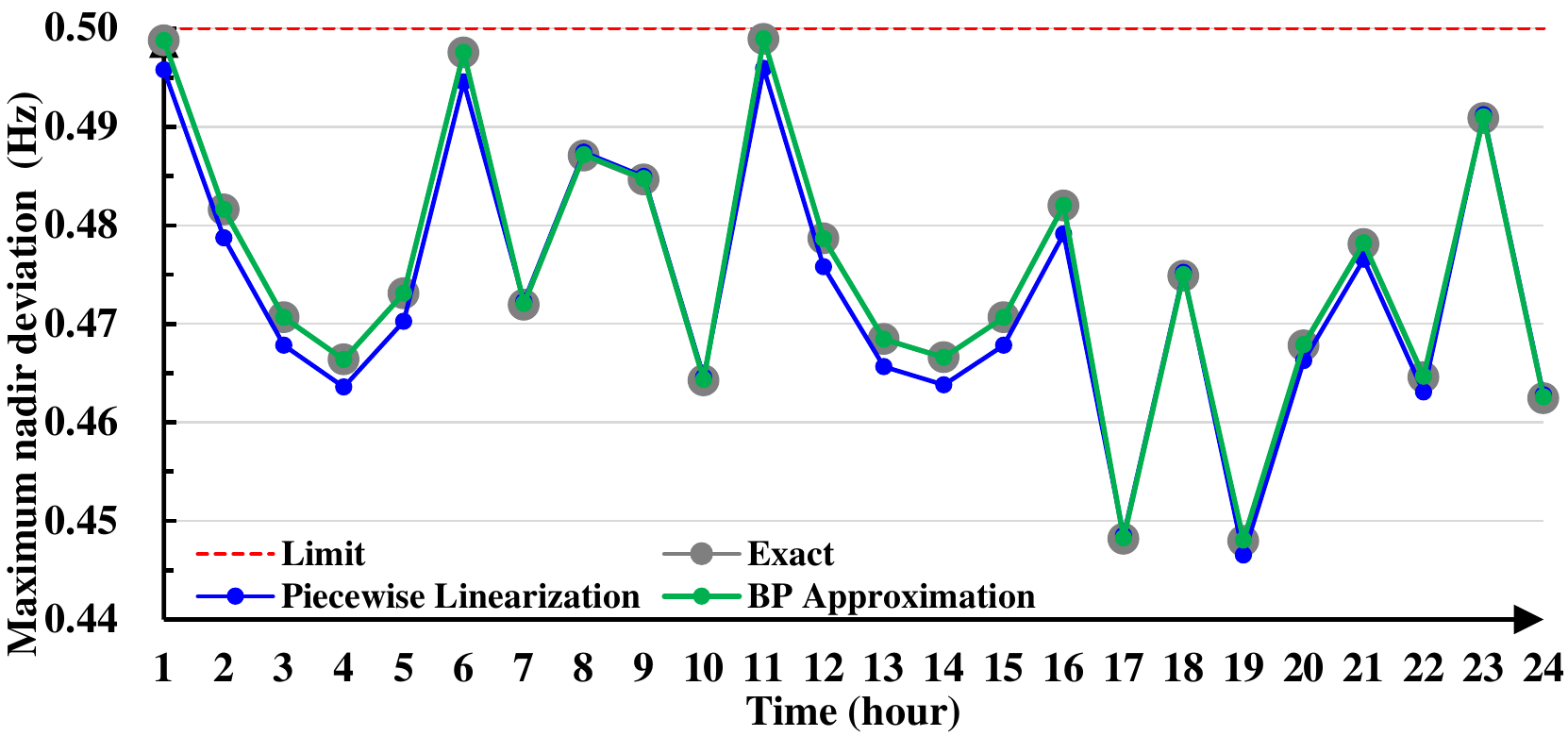}
                \vspace{-1ex}
                \caption{Nadir frequency deviation in Case 5}	
                \vspace{-1ex}
                \label{fig:nadir}
            \end{figure}

            We report the performance details in TABLE \ref{tab:results5}, and the optimal UC in Case 5 is the same as in Fig. \ref{fig:UC3}.
            We can see that the total cost of Case 5 is higher than that of Case 3, mainly due to the conservative PFR reserve of units.
            The computational time of Case 5 is much less than that of Case 3, because the piecewise linearization method induces a more tractable nadir constraint.
            With the solution of Case 5, we compare the exact nadir deviation and the estimated nadir deviation of piecewise linearization and BP approximation in each period, and the results are shown in Fig. \ref{fig:nadir}.
            We observe that BP approximation has higher accuracy in estimating nadir deviation.
            In addition, piecewise linearization usually provides a lower estimation than the exact value of the nadir deviation (see Fig. \ref{fig:nadir}), which is optimistic and may cause security risks.

        \subsubsection{Analysis on DRCCs}
            We first compare the deterministic approach and the adopted DRCC approach (Case 2).
            The deterministic approach gives the same UC solution as in Case 2 and incurs a lower operation cost.
            Nevertheless, the DRCC approach explicitly considers the possibility of having less wind power available than predicted and allocates more generation and reserves on the thermal units, which helps reduce the re-dispatch of thermal units \cite{ZhouFang-2644}.
            The re-dispatch cost coefficient is set as 40\$/MWh and 1000 scenarios are randomly generated.
            So the expected re-dispatch costs of the deterministic approach and the DRCC approach are 819.01\$ and 45.76\$, respectively.
            The latter is far less than the former, which validates the effectiveness of DRCC in reducing risks.

            \begin{table}[!t]
                \renewcommand{\arraystretch}{1.0}
                \vspace{-0ex}
                \caption{Detailed Comparison Between Cases 3 and 5}
                \vspace{-2ex}
                \label{tab:results5}
                \centering
                \newcommand{\tabincell}[2]{\begin{tabular}{@{}#1@{}}#2\end{tabular}}
                \begin{tabular}{ccc}
                    \toprule
                    Comparison terms & Case 3 & Case 5\\
                    \midrule
                    Start-up \& shut-down cost (\$) & 0 & 0\\
                    Fuel cost of thermal units (\$) & 47718.70 & 47878.25\\
                    PFR reserve cost of thermal units (\$) & 4712.09 & 7740\\
                    PFR reserve cost of wind farm (\$) & 1033.06 & 1100\\
                    Total cost (\$) & 53463.85 & 56718.25\\
                    \midrule
                    Maximum RoCoF (Hz/s) & 0.2354 & 0.2354\\
                    Maximum $\Delta f$ at the nadir (Hz) & 0.4518 & 0.4959\\
                    Maximum $\Delta f$ at the quasi-steady state (Hz) & 0.2996 & 0.2996\\
                    \midrule
                    Computation time (s) & 57.15 & 14.20\\
                    \bottomrule
                \end{tabular}
                \vspace{-0ex}
            \end{table}
            \begin{figure}[!t]
                \centering
                \subfigure[Risk level $\epsilon$]{\includegraphics[width=0.45\textwidth]{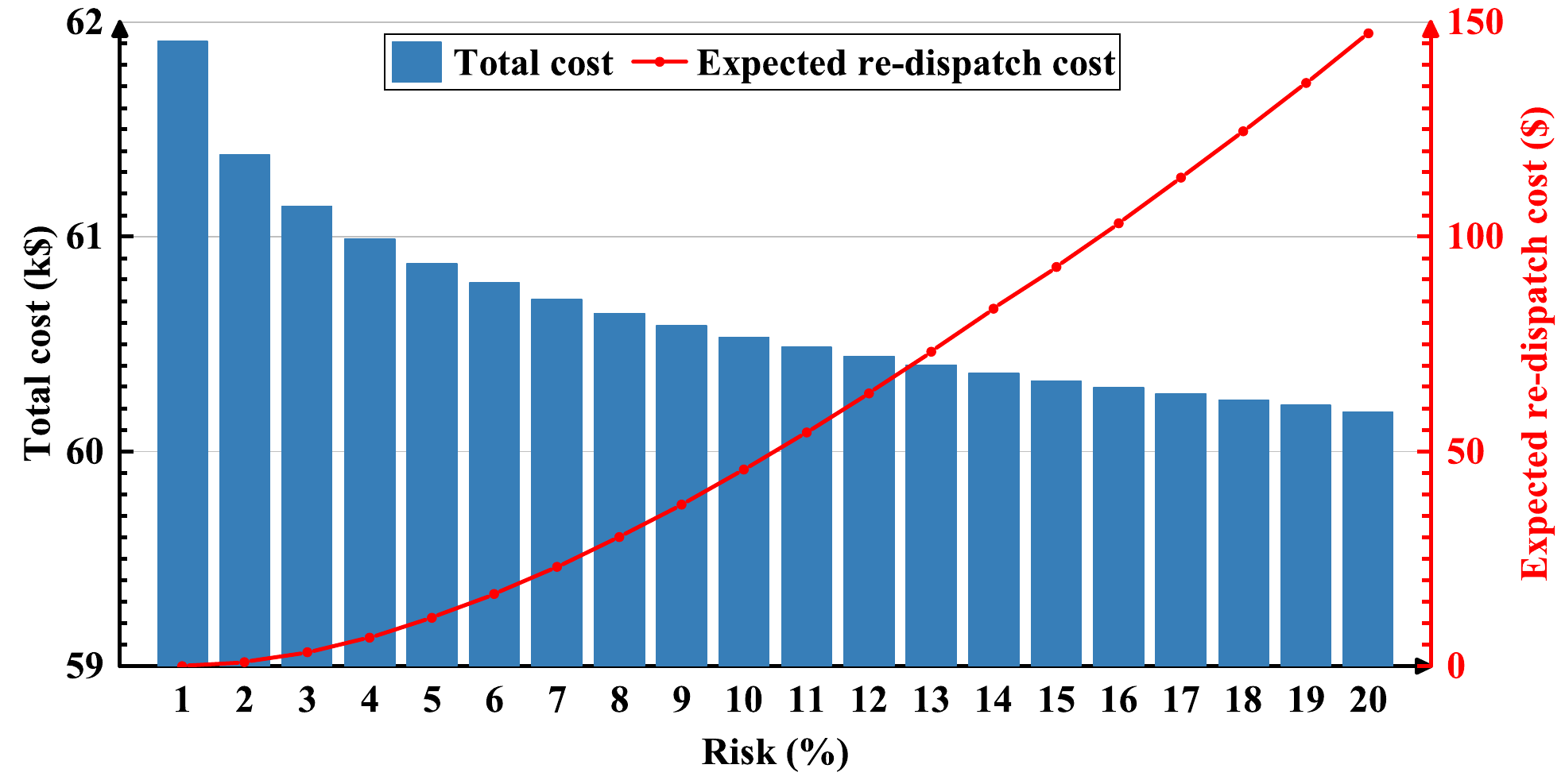} }\\
                \subfigure[Radius $r$ of the Wasserstein ball]{\includegraphics[width=0.45\textwidth]{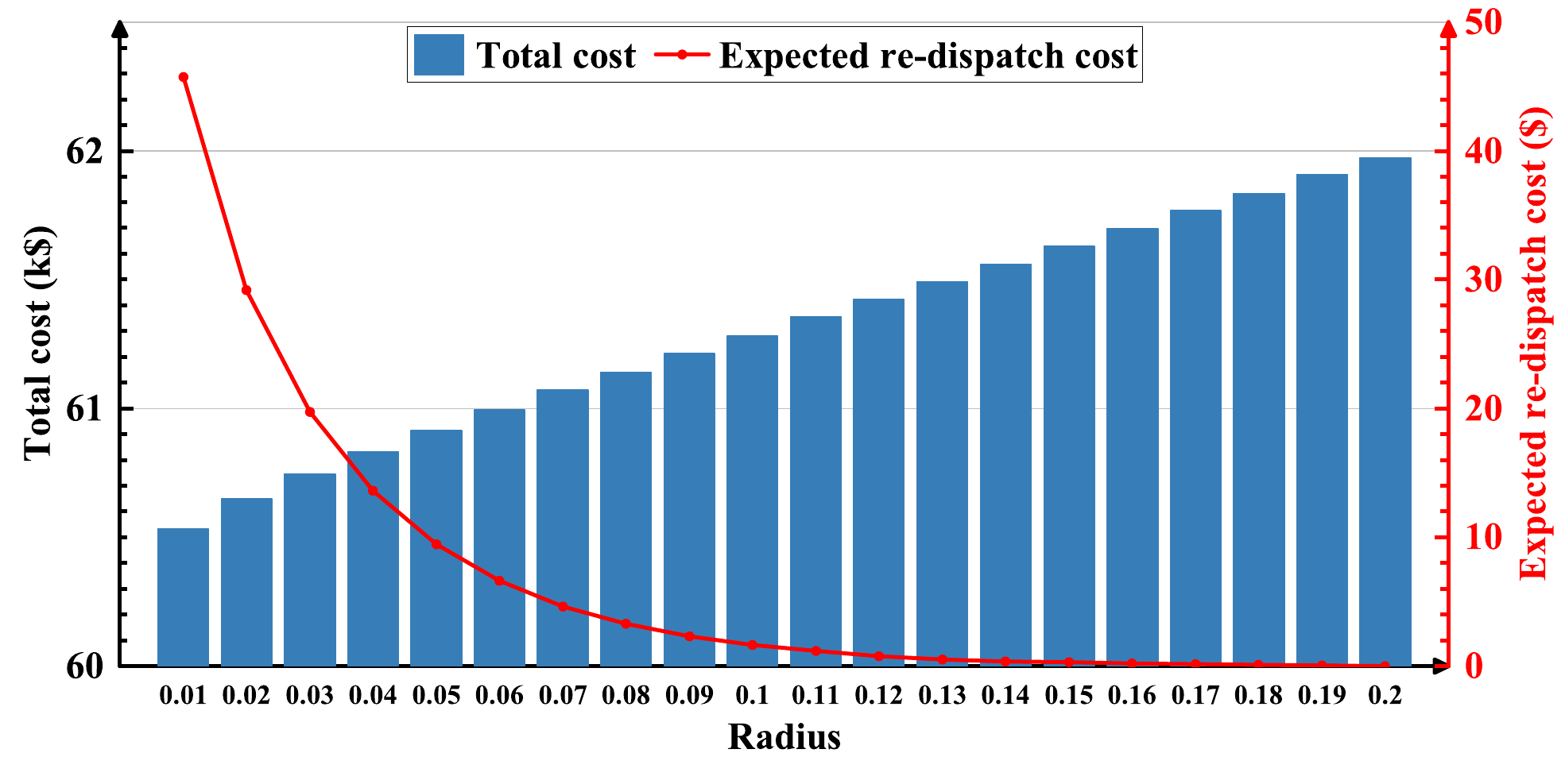} }\\
                \vspace{-1ex}
                \caption{DRCC sensitivity analysis}
                \vspace{-2ex}
                \label{fig:DRCC}
            \end{figure}

            Next, we vary the risk level $\epsilon$ and the Wasserstein ball radius $r$ to see their impacts on the total cost of the UC model and the expected re-dispatch cost.
            The results are reported in Fig.~\ref{fig:DRCC}.
            This figure provides clear trade-offs among operational economy, risk attitude, and robustness, which can better assist the system operators.
            In particular, the total cost is more sensitive to the risk level than it is to the Wasserstein ball radius, especially when $\epsilon$ is small (or equivalently, when $1 - \epsilon$ is close to $1$).
            When $\epsilon$ exceeds 0.1, the sensitivity of the total cost to $\epsilon$ and $r$ are similar.

        \subsection{Scalability Test in IEEE 118-Bus System}
            We evaluate the scalability of our approach using the IEEE 118-bus system, which includes 54 thermal units and a 4242 MW of maximum total load.
            We follow~\cite{YangXu-3622} on the frequency-related parameters.
            We scale up the capacity of the wind farm to 800 MW and accordingly the wind power and the range of the droop factors.
            The results are reported in Table~\ref{tab:results118}.
            \begin{table}[!htbp]
                \renewcommand{\arraystretch}{1.0}
                \vspace{-0ex}
                \caption{Results in IEEE 118-Bus System}
                \vspace{-2ex}
                \label{tab:results118}
                \centering
                \newcommand{\tabincell}[2]{\begin{tabular}{@{}#1@{}}#2\end{tabular}}
                \begin{tabular}{ccc}
                    \toprule
                    Comparison terms & Case 1 & Case 2\\
                    \midrule
                    Operation cost of thermal units (k\$) & 2184.18 & 2394.85\\
                    PFR reserve cost of thermal units (k\$) & 130.21 & 155.65\\
                    PFR reserve cost of wind farm (k\$) & 0 & 9.81\\
                    Total cost (k\$) & 2314.39 & 2560.31\\
                    \midrule
                    Maximum RoCoF (Hz/s) & 0.2598 & 0.1696\\
                    Maximum $\Delta f$ at the nadir (Hz) & 0.7748 & 0.4318\\
                    Maximum $\Delta f$ at the quasi-steady state (Hz) & 0.4950 & 0.1935\\
                    \midrule
                    Computation time (s) & 49.98 & 2472\\
                    \bottomrule
                \end{tabular}
                \vspace{-0ex}
            \end{table}

\begin{figure}[!htbp]	
                \centering
                \vspace{-0ex}
                \includegraphics[width=0.45\textwidth]{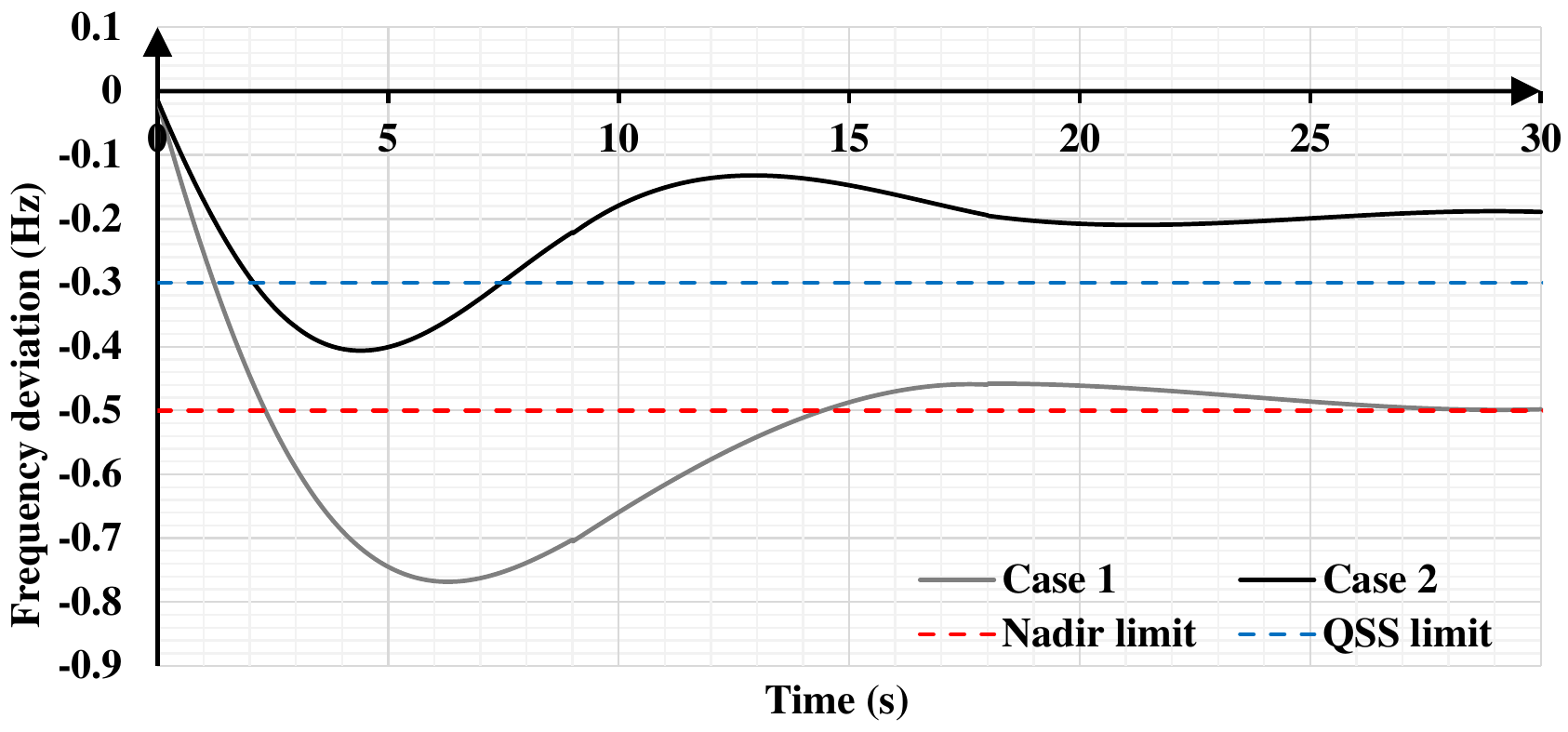}
                \vspace{-1ex}
                \caption{Comparison of frequency dynamics in IEEE 118-bus system}	
                \vspace{-0ex}
                \label{fig:dynamics-case118}
            \end{figure}

\begin{figure}[!htbp]	
                \centering
                \vspace{-0ex}
                \includegraphics[width=0.45\textwidth]{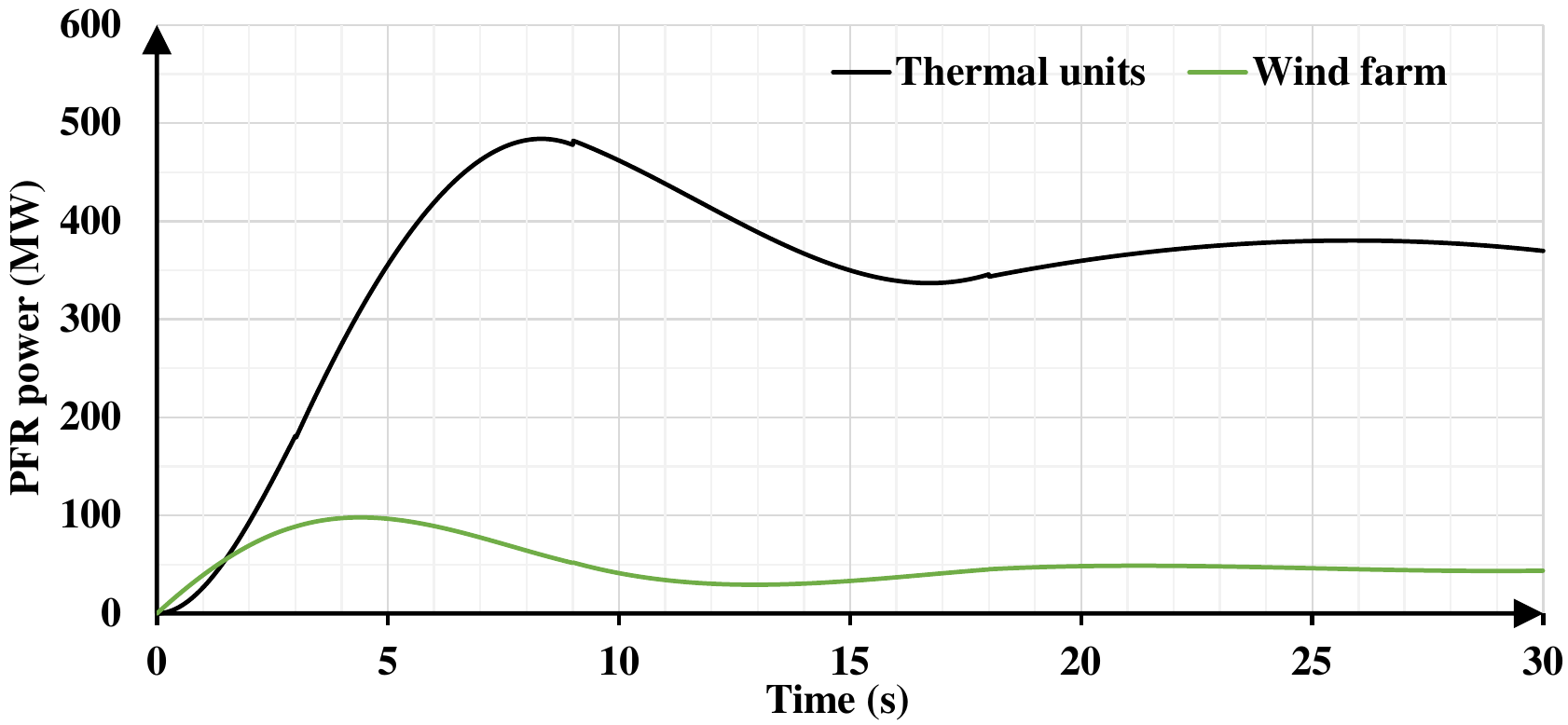}
                \vspace{-1ex}
                \caption{PFR power dynamics of Case 2 in IEEE 118-bus system}	
                \vspace{-0ex}
                \label{fig:PFR-case118}
            \end{figure}

            From Table~\ref{tab:results118}, we observe that in Case 1, the frequency deviation of hour 21 reaches 0.7748 Hz at the nadir and reaches 0.4950 Hz at the QSS, exceeding the frequency security limits.
            In contrast, in Case 2, all frequency security limits are respected. (see Fig.~\ref{fig:dynamics-case118}).
            Nevertheless, the computational time in Case 2 is much longer than that in Case 1, which can be seen as the cost of guaranteeing frequency security.
            It is hence interesting to further improve the computational tractability of the proposed method and investigate the right trade-off between its performance and computational burden in future study.
            Fig.~\ref{fig:PFR-case118} provides the PFR power in Case 2.
            These results demonstrate the effectiveness and scalability of the proposed method in protecting frequency security in larger-sized systems.


    \vspace{-0ex}
    \section{Conclusions}\label{sec:conlcusion}
        This paper proposed a frequency stability-constrained UC model considering frequency dynamics, wind power uncertainty, and variable droop factors.
        In this model, we incorporated frequency dynamics using DAEs and considered the dead band.
        Then, we derived an inner approximation of the DAEs using BP splines and linearization techniques.
        A comparison with Simulink demonstrated the tightness of the BP approximation in depicting frequency dynamics and evaluating the frequency deviation at the nadir.
        Extensive case studies based on IEEE 6-bus and 118-bus systems demonstrated the effectiveness of the proposed model in deciding UC and PFR reserves for frequency security, the necessity of considering the dead band in frequency dynamics, and the value of using variable droop factors of the wind farms in improving the coordination among different frequency support resources, which can help reduce the total cost significantly.
        Future works will consider more types of frequency response resources, such as energy storage, to support frequency stability in a market environment.

    \bibliographystyle{IEEEtran}
    \bibliography{IEEEabrv,BibTexRef}

\end{document}